\documentclass[aps,pre,groupedaddress,showpacs,preprint,longbibliography]{revtex4-1} 

\usepackage{graphicx}
\usepackage{bm}
\usepackage{amssymb}
\usepackage{amsmath}
\usepackage{color,soul}
\usepackage{ulem}
\usepackage{comment} 

\usepackage{color,soul}  
\setstcolor{red}         

\newcommand{\be}{\begin{equation}}
\newcommand{\ee}{\end{equation}}
\newcommand{\ba}{\begin{eqnarray}}
\newcommand{\ea}{\end{eqnarray}}
\newcommand{\besu}{\begin{subequations}}
\newcommand{\esu}{\end{subequations}}

\begin{document}
\title{The standard and the fractional Ornstein-Uhlenbeck process on a growing domain}
\author{F. Le Vot$^{1}$, S.~B. Yuste$^{1}$, and E. Abad$^{2}$}
\affiliation{
 $^{1}$ Departamento de F\'{\i}sica and Instituto de Computaci\'on Cient\'{\i}fica Avanzada (ICCAEx) \\
 Universidad de Extremadura, E-06071 Badajoz, Spain \\
 $^{2}$ Departamento de F\'{\i}sica Aplicada and Instituto de Computaci\'on Cient\'{\i}fica Avanzada (ICCAEx) \\ Centro Universitario de M\'erida \\ Universidad de Extremadura, E-06800 M\'erida, Spain
}

\begin{abstract}
We study normal diffusive and subdiffusive processes in a harmonic potential (Ornstein-Uhlenbeck process) on a uniformly growing/contracting domain. Our starting point is a recently derived fractional Fokker-Planck equation, which covers both the case of Brownian diffusion and the case of a subdiffusive Continuous-Time Random Walk (CTRW). We find a high sensitivity of the random walk properties to the details of the domain growth rate, which gives rise to a variety of regimes with extremely different behaviors. At the origin of this rich phenomenology is the fact that the walkers still move while they wait to jump, since they are dragged by the deterministic drift arising from the domain growth. Thus, the increasingly long waiting times associated with the ageing of the subdiffusive CTRW imply that, in the time interval between two consecutive jumps, the walkers might travel over much longer distances than in the normal diffusive case. This gives rise to seemingly counterintuitive effects. For example, on a static domain, both Brownian diffusion and subdiffusive CTRWs yield a stationary particle distribution with finite width when a harmonic potential is at play, thus indicating a confinement of the diffusing particle. However, for a sufficiently fast growing/contracting domain, this qualitative behavior breaks down, and differences between the Brownian case and the subdiffusive case are found. In the case of Brownian particles, a sufficiently fast exponential domain growth is needed to break the confinement induced by the harmonic force; in contrast, for subdiffusive particles such a breakdown may already take place for a sufficiently fast power-law domain growth. Our analytic and numerical results for both types of diffusion are fully confirmed by random walk simulations.
\end{abstract}


\maketitle


\section{Introduction}

Diffusion in growing domains is a phenomenon of great interest in a number of disciplines, e.g., in cosmology
\cite{Berezinsky2006, Berezinsky2007, Lagutin2003, Uchaikin2013} and in biology \cite{Murray2003}. In cosmology, examples include scattering of electromagnetic radiation in inhomogeneous magnetic fields \cite{Kotera, Batista2014} during the expansion of the universe; in biological media, particle dissemination often takes place on time scales where the physical growth of the embedding medium cannot be neglected. For instance, embryonic tissue growth via cellular division takes place during the spreading process leading to the formation of a morphogen gradient \cite{Fried2014, Averbukh2014}, whereby the local concentration of morphogens may influence the growth process itself \cite{Averbukh2014}.  Diffusion of substances throughout growing organs \cite{Murray2003, Kulesa1996, Crampin1999, Crampin2001, Landman2003, Binder2008, Baker2010, Simpson2015a, Simpson2015b, Simpson2015c, Ross2016} has also been invoked to explain phenomena such as the formation of pigmentation patterns \cite{Murray2003}, teeth primordia in animals \cite{Kulesa1996}, or the growth of microorganisms into colonies \cite{Vulin2014}. Finally, stochastic transport in growing domains is also a topic of great interest in finance, where random walk models have played a major role since the seminal work of Bachelier \cite{Bachelier1900}. In this context, the phenomenon of inflation \cite{Ahmad08} can be thought of as an additional shift in a walker's position arising from a dilation of the spatial domain.

Even though diffusion in growing domains has been studied over several decades, a surprisingly large number of open questions remain. For example, despite recent progress \cite{YAE2016, AmasApaper, OurBookChapter1, OurBookChapter2}, a comprehensive theory of first-passage processes and encounter-controlled reactions in growing domains is still missing. Another open problem is the characterization of biased random walks in growing domains. Of special interest in this context is the effect of anomalous diffusion processes, which have attracted great interest in recent years as a means to model stochastic transport in biological media \cite{LubKlaft08, Magdziarz09a, SokoReview12, Hofling2013, MetzlerJeon14, MerozSoko15, Tan2018, Etoc18}. The interplay between the anomalous diffusion process and the domain growth may give rise to nontrivial effects, both in the absence \cite{LeVotAbadYuste17, Angstmann17} and in the presence \cite{LeVotYuste2018} of a biasing force. Recent work has e.g. shown that biased walks on growing domains violate the generalized Einstein relation
\cite{LeVotYuste2018}, even in the case of normal diffusion. While interesting effects already arise in the case of a constant force \cite{LeVotYuste2018}, a more complex type of bias, e.g., one arising from a Hookean force, is expected to yield an even richer phenomenology in combination with anomalous diffusion processes. Exploring this phenomenology is the main goal of this paper, both for normal and for anomalous diffusion.

 On a fixed domain, the combination of normal diffusion and the harmonic potential associated with the Hookean force yields the celebrated Ornstein-Uhlenbeck (OU) process \cite{OU30}.  Indeed, the OU process and variants thereof \cite{Black73, Merton73, Vasicek77, Magdziarz05, Magdziarz09b, Grebenkov14, Onalan2015, Chen2019} have found a wide variety of applications in different fields such as Finance \cite{Black73, Merton73, Vasicek77}, Optics \cite{Grebenkov14}, etc. In particular, subdiffusive versions of the OU process are relevant for the field of financial mathematics \cite{Onalan2015}. Here, we incorporate an additional ingredient to the OU process, namely, a kinetic process which results in the growth or contraction of the spatial domain in which the stochastic transport takes place.

 The remainder of this paper is organized as follows. In Sec.~\ref{Sec:OU_FFPE}, we briefly outline the derivation of a recently obtained fractional Fokker-Planck equation (FFPE) \cite{LeVotYuste2018} that is then taken as a starting point for subsequent calculations. Secs.~\ref{Sec:OU_Normal} and~\ref{Sec:OU_SubDiff} are respectively devoted to the phenomenology of the normal diffusive and of the subdiffusive process on a uniformly growing/contracting domain. Finally, in Sec.~\ref{Sec_Conclusions}, we summarize our main conclusions and suggest possible ways of extending the present work.

\section{CTRW model and FFPE for diffusion on a growing domain}
\label{Sec:OU_FFPE}

In Ref.~\cite{LeVotYuste2018}, a FFPE for a subdiffusive random walk evolving on a uniformly growing domain was derived. The starting point to obtain the FFPE is a Continuous-Time Random Walk (CTRW) model \cite{MontrollWeiss, Metzler00} in which the diffusive particles (also called walkers in what follows) are subject to the action of an external force. The walkers are assumed to perform
instantaneous jumps at randomly distributed times. The force field is only at play when the particles jump, in which case it induces a bias in the jump direction. In the separable version of the model, the statistics of the walk is dictated by the waiting time and the jump length probability distribution functions (pdfs).

To set the stage, we will focus on the $1d$ case (a generalization to the higher-dimensional case proceeds along similar lines). A given physical point on the $1d$ domain is shifted as a result of the domain growth. Thus, its coordinate $y$ changes in the course of time. For convenience, one also defines a so-called comoving coordinate $x$, which is simply the initial position $y_0$ of this point. For a uniformly growing domain, one has a simple relationship between the physical coordinate and the comoving coordinate, namely,
\begin{equation}
y(t) = a(t) x.
\label{y=a(t)x}
\end{equation}
In the language of cosmology, the quantity $a(t)$ is called the ``scale factor'' \cite{Ryden2003}. For a growing domain, one has $a(t)>1$. Note, however, that the formalism introduced hereafter also accounts for the case $0<a(t)<1$ of a contracting domain. In the sequel, we shall occasionally use the terms ``expanding'' and ``shrinking'' as synonyms of ``growing'' and ``contracting''.

Let us assume that $y$ describes the position of a walker performing a biased, separable
CTRW on the growing domain. Two different types of biasing forces can be distinguished, namely, those acting on the walker at all times \cite{Metzler00, Cairoli18, Chen2019}, and those that are only at play when the walker jumps, resulting in an asymmetric jump length pdf \cite{Metzler00, Metzler1998, LeVotYuste2018, Chen2019}. As already mentioned, we shall focus on the second case here. Were the walk taking place on a static domain, the particle would remain at the same physical position between two consecutive jumps; on a growing or shrinking domain, this is no longer true, since the particle is drifted in physical space because of the displacement of the (expanding/contracting) ``volume element'' in which it dwells. From the point of view of the walker's motion, this kind of drift can be viewed as arising from a physical force; in fact, the effect of an exponential contraction on the diffusing particle turns out to be equivalent to the action of the harmonic force in the OU process \cite{Chen2019}.

In our model, the effect of the force field will be included by means of the following jump length distribution \cite{LeVotYuste2018}:
\begin{equation}
\Lambda^{*}(y,y',t) = 2 \lambda^{*}(y-y') \left[ A^{*}(y',t) \Theta(y-y') + B^{*}(y',t) \Theta(y'-y)  \right],
\label{LambdaChoice}
\end{equation}
where $\lambda^{*} (y-y')$ is a symmetric pdf for the jump length $|y-y'|$ with a typical variance $2 \sigma^2$.
In Eq.~\eqref{LambdaChoice}, $A^{*}(y',t)$ [$B^{*}(y',t)$] denotes the probability that a random walker located at $y'$ takes an instantaneous jump to the right (left) at time $t$. Obviously, one has $B^{*}(y',t)=1-A^{*}(y',t)$.

In the comoving reference frame, the corresponding jump length pdf $\Lambda(x,x',t)$ can be derived by applying probability conservation arguments. One has, $\Lambda^{*}(y,y') dy = \Lambda(x,x',t) dx$. Hence,
\begin{equation}
\Lambda(x,x',t) = a(t) \Lambda^{*}(a(t) x, a(t) x', t) .
\end{equation}
This can be written out as follows:
\begin{equation}
\Lambda(x,x',t) = 2 \lambda(x,x',t) \left[ A(x',t) \Theta(x-x') \right. + \left. B(x',t)  \Theta(x'-x) \right],
\end{equation}
where $A(x,t)=A^{*}(a(t) x,t)$, $B(x,t)=B^{*}(a(t) x',t)$ and $\lambda(x,x',t) = a(t) \lambda^{*}(a(t) (x- x'))$.
As one can see, in the comoving reference frame the pdf $\lambda$ adopts the same form as the pdf corresponding to a CTRW process on a static domain, except for the fact that a time dependence comes in. This time dependence arises from the change of the jump length by a factor $1/a(t)$ in the comoving reference frame.

In what follows, we will consider two special yet important types of pdfs. The first class has the following long-time asimptotic form:
\begin{equation}
\varphi (\Delta t) \sim \frac{\alpha \tau^{\alpha} }{\Gamma (1-\alpha)} (\Delta t)^{-1-\alpha}
\label{heavytailwtd}
\end{equation}
with $0<\alpha<1$ and the characteristic time $\tau$. In the standard case of a static domain, this type of pdf is known to yield subdiffusion, i.e., diffusion with a sublinear growth of the mean square displacement (msd) $\propto t^\alpha$. In particular, for our numerical random walk simulations we will use the Pareto pdf:
\begin{equation}
\varphi (\Delta t) =\frac{\alpha}{\omega_{\alpha}} \left( 1 + \frac{\Delta t}{\omega_{\alpha} } \right)^{-1-\alpha},
\end{equation} where $\omega_{\alpha} = \tau / [\Gamma (1- \alpha )]^{1/\alpha}$.
The other class of pdfs we will consider are those leading to normal diffusion (msd $\propto t$).  In particular we will use
the exponential distribution:
\begin{equation}
\varphi(\Delta t) = \frac{1}{\tau} \exp \left( -\frac{t}{\tau} \right),
\label{Exponential PDF}
\end{equation}
although any other pdf with a finite first-order moment would also yield normal diffusion.

Let us now introduce the pdf $W(x,t)$, which is associated with the (infinitesimal) probability $W(x,t) dx$ of finding the walker within the interval delimited by $x$ and $x + dx$ at time $t$. This pdf has of course a counterpart in physical space, hereafter denoted by $W^{*}(y,t)$. From probability conservation, one has $W^{*}(y,t)=W(y/a(t),t)/a(t)$. The FFPE for $W(x,t)$ is \cite{LeVotYuste2018}:
\begin{equation}
\frac{\partial W(x,t)}{\partial t} = \frac{\mathfrak{D}_{\alpha}}{a^2(t)} {_0}\mathcal{D}^{1-\alpha}_t \left[ \frac{\partial^2 W}{\partial x^2} \right]
- \frac{1}{a(t)} \frac{1}{\xi_{\alpha}} \frac{\partial}{\partial x} \left[ F(x,t) ~ {_0}\mathcal{D}^{1-\alpha}_t W(x,t) \right]
\label{EDADME_Force}
\end{equation}
with $F(x,t)=F^{*}(a(t) x, t)$ and the anomalous diffusion coefficient
\begin{equation}
\mathfrak{D}_{\alpha} = \frac{\sigma^2}{\tau^{\alpha}}.
\label{DiffusionConstant}
\end{equation}
In Eq.~\eqref{EDADME_Force},
\begin{equation}
\xi_{\alpha} = \frac{F^{*} \tau^{\alpha}}{2(A^{*}-B^{*}) \varepsilon \sigma}
\label{FrictionConstant}
\end{equation}
stands for the generalized friction constant, where
\begin{equation}
\varepsilon = \frac{1}{\sigma} \int_0^{\infty} dy ~ y \lambda^{*}(y) ,
\label{defeps}
\end{equation}
is a dimensionless constant. Taken together with the constraint $|A^{*}-B^{*}|\le 1$, Eq.~\eqref{FrictionConstant} implies that the effect of the force saturates if its magnitude exceeds the limit $2\xi_{\alpha} \varepsilon \sigma / \tau^{\alpha}$; as soon as this is the case, the bias reaches its maximum value, and so the particle always jumps in the same direction (to the right, say, if $A^*=1$ and $B^*=0$). The integro-differential operator on the right hand side (rhs) of~\eqref{EDADME_Force} is the so-called Gr\"unwald-Letnikov fractional derivative of order $1-\alpha$. A straightforward definition of this operator in terms of the Laplace transform of its argument is the following:
\begin{equation}
{_0}\mathcal{D}^{1-\alpha}_t f(t) = \mathcal{L}^{-1} \left[ s^{1-\alpha} \tilde{f}(s) \right],
\end{equation}
where the Laplace transform  $\mathcal{L}$ is defined by
\begin{equation}
\mathcal{L} [f](s) = \tilde{f}(s) = \int_0^{\infty} dt f(t) \exp(-st) .
\end{equation}
When the function $f(t)$ is continuous and sufficiently well-behaved at the origin (see, e.g., Eqs.~(2.255), (2.248) and (2.240)
in Ref.~\cite{Podlubny1999}), the Gr\"unwald-Letnikov operator is equivalent to the Riemann-Liouville fractional derivative, defined as follows:
\begin{equation}
  {_{~~0}^{RL}}D^{1-\alpha}_t f(t) = \frac{1}{\Gamma(\alpha)} \frac{\partial}{\partial t} \int_0^t du \frac{f(u)}{(t-u)^{1-\alpha}} .
\end{equation}
In the OU process, the diffusive particles are subjected to an external harmonic potential associated with the Hookean force $F^{*}(y)=-\kappa y$, which is directed towards the potential minimum at $y=0$. In terms of the comoving coordinate, one has $F(x,t) = -\kappa x a(t)$, and consequently, Eq.~\eqref{EDADME_Force} takes the following form:
\begin{equation}
\frac{\partial W(x,t)}{\partial t} = \frac{\mathfrak{D}_{\alpha}}{a^2(t)} {_0}\mathcal{D}^{1-\alpha}_t \left[ \frac{\partial^2 W}{\partial x^2} \right] + \frac{\kappa}{\xi_ \alpha} \frac{\partial}{\partial x} \left[ x ~ {_0}\mathcal{D}^{1-\alpha}_t W(x,t) \right] ,
\label{FPE_Harmonic}
\end{equation}
Equation~\eqref{FPE_Harmonic} will be the starting point of our subsequent analysis. The resulting solutions will be compared with the outcome of random walk simulations (see Appendix).

\section{OU process for a Brownian particle on a growing domain}
\label{Sec:OU_Normal}

In the case of a Brownian particle ($\alpha=1$), Eq.~\eqref{FPE_Harmonic} takes the simplified form
\begin{equation}
\frac{\partial W(x,t)}{\partial t} = \frac{\mathfrak{D}}{a^2(t)} \frac{\partial^2 W}{\partial x^2} +
\frac{\kappa}{\xi} \frac{\partial}{\partial x} \left[ x W(x,t) \right] ,
\label{OUEq_Comoving}
\end{equation}
with $\mathfrak{D} \equiv \mathfrak{D}_1$ and $\xi\equiv \xi_1$. It is worth noting that Eq.~\eqref{OUEq_Comoving} may not only be derived in the framework of the CTRW model, but also by means of other methods, such as the Langevin approach introduced in Refs.~\cite{YAE2016} and \cite{OurBookChapter1}. The corresponding Langevin equation in physical coordinates reads
\begin{equation}
y(t+dt)=y(t)+(v_H+v)dt+\sqrt{2\mathfrak{D}}\,d{\cal W}(t),
\label{Leq}
\end{equation}
where ${\cal W}(t)$ stands for a Wiener process, $v=F^{*}(y)/\xi=-\kappa y/\xi $ is the intrinsic drift velocity, and $v_H = y \dot{a}/a $ stands for the drift velocity associated with the deterministic domain growth (in the language of cosmology, $v_H$ is the ``Hubble velocity'').

As in the case of a static domain, the propagator solution of Eq.~\eqref{OUEq_Comoving}, i.e., the solution corresponding to the initial condition $W(x,0)=\delta(x-x_0)$ on the infinite 1$d$ line, can be obtained by the method of characteristics (see e.g. Sec. 3.8.4 in Ref.~\cite{Gardiner2010}). One finds
\begin{equation}
\widehat{W} (k,t) = \widehat{W}_0 (k(t)) \exp \left[- k^2 \sigma_x^2 (t) \right],
\label{fteq}
\end{equation}
with $k(t)=k_0 \exp(\kappa t /\xi)$ and
\begin{equation}
\sigma_x^2 (t) = \mathfrak{D} \exp(-2 \kappa t/\xi) \int_0^t du \, \frac{\exp(2 \kappa u /\xi)}{a^2 (u) }.
\label{sigmaOU}
\end{equation}
The hat symbol in Eq.~\eqref{fteq} denotes the Fourier transform, defined as follows:
\begin{equation}
\mathcal{F} [f(x)]= \widehat{f} (k) = \int_{-\infty}^{\infty} dx \exp(-i k x) f(x).
\end{equation}
At the initial time, $t=0$, the walker is assumed to be at the position $x_0$. This implies
\begin{equation}
\widehat{W}_0 (k_0) = \exp(i k_0 x_0) = \exp \left[ i k \exp \left(- \kappa t/\xi \right) x_0 \right].
\end{equation}
The inverse Fourier transform of the rhs is a Gaussian distribution
\begin{equation}
W(x,t) = \frac{1}{\sqrt{4 \pi \sigma_x^2 (t)}} \exp \left( -\frac{ \left[x- \langle x (t) \rangle \right]^2}{4 \sigma_x^2} \right)
\label{comprop},
\end{equation}
with the (time decaying) mean value
\begin{equation}
\langle x (t) \rangle = \exp \left( - \kappa t / \xi \right)=x_0 \exp \left( - t / t_r \right),
\label{averageOU}
\end{equation}
where the characteristic relaxation time $t_r\equiv \xi/\kappa$ has been introduced. The expression for $t_r$ tells us that a small friction or a large harmonic force favor a quick localization of the particle about the origin. As one might have anticipated, the behavior described by Eq.~\eqref{averageOU} turns out to be independent of the scale factor $a(t)$.

From Eq.~\eqref{comprop}, the propagator in physical space (also termed ``physical propagator'' hereafter) immediately follows as
\begin{equation}
W^{*}(y,t) = \frac{1}{\sqrt{4 \pi \sigma_y^2 (t)}} \exp \left( -\frac{ \left[ y- \langle y (t) \rangle \right]^2}{4 \sigma_y^2(t)} \right),
\label{GaussProp}
\end{equation}
with $\langle y (t) \rangle = a(t) \langle x (t) \rangle $ and $\sigma_y^2 (t) = a^2(t) \sigma_x^2 (t)$. Note that the above result contains as a special case the solution corresponding to a static domain  ($a(t)\equiv 1$). This special case is characterized by a time-dependent semi-variance $\sigma_y^2(t)$, namely,
\begin{equation}
\sigma_y^2 (t) =  \frac{1-\exp (-2 t/t_r)}{2}\,\mathfrak{D} t_r.
\label{s2yOEestatico}
\end{equation}
When $y_0\equiv x_0=0$, the propagator approaches an equilibrium Gaussian distribution whose semi-variance is $\sigma^2_y (\infty)=\mathfrak{D} t_r/2$, which means that the width of the particle distribution eventually stabilizes as a result of the trade-off between diffusive spreading and the strong localization induced by the harmonic force.

What happens in the case of a growing/contracting domain? Here, the domain growth exerts a drag force on the particle which results in a deterministic drift (``Hubble drift''). The Hubble drift will be directed towards the origin if the domain shrinks, or away from it if the domain grows. One may easily guess that the behavior of the system will be very sensitive to the functional form of the scale factor. This is indeed confirmed by a detailed analysis, which is carried out in the next two subsections for the special cases of a power-law scale factor and of an exponential scale factor.

\subsection{Power-law growth}
\begin{figure}[t]
\includegraphics[width=0.49\textwidth]{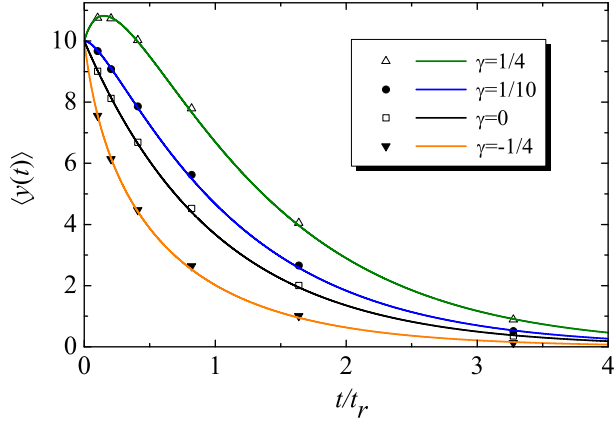}
\caption{
Time evolution of $ \langle y \rangle $ for $y_0=10$, $\mathfrak{D}=1/2$ and $t_r=\xi/\kappa=10^4$. Solid lines depict the theoretical solutions for a power-law expansion -see Eq.~\eqref{ybarPot}- with $t_0=10^3$ and $\gamma=\{1/4,1/10,0,-1/4\}$, from top to bottom. The different symbols correspond to simulation results.}
\label{Fig:MeanValueHarmonicPL}
\end{figure}

Our specific goal here will be to explore the behavior of $W^{*}(y,t)$ for a scale factor of the form $a(t)=[(t+t_0)/t_0]^\gamma$. As we have seen, the solution for a deterministic initial condition of the form $\delta(y-y_0)$ is a Gaussian bell with a time-dependent mean value and variance.  Since the relation $\langle y \rangle = a(t) \langle x \rangle$ holds, the first-order moment of the physical coordinate is obtained by multiplying Eq.~\eqref{averageOU} with the scale factor $a(t)$.  The resulting expression is
\begin{equation}
\langle y \rangle = y_0 \exp \left(- \frac{t}{t_r} \right) \left( \frac{t+t_0}{t_0} \right)^{\gamma}.
\label{ybarPot}
\end{equation}
Since the constant $t_r$ is always positive, $\langle y \rangle$ tends to 0 (the minimum of the harmonic potential) at long times. However, the transient behavior depends on the sign of the expression $v_H+v=\dot{a}/a-1/t_r$, which appears in the first term on the rhs of the Langevin equation~\eqref{Leq}. For  a power-law scale factor, one has $v_H+v=\gamma/(t+t_0)-1/t_r$. Thus, for $t_0 \ge \gamma t_r$, one has $v_H+v<0$ at any time $t>0$; this implies that the decay is monotonic, reflecting the fact that at all times the frictional force is small enough and the harmonic potential sufficiently stiff for the walker to overcome the Hubble drift. In contrast, when $t_0 < \gamma t_r$, the decay to the origin is non-monotonic; the particle tends to move away from $y=0$ as long as $v_H+v>0$, i.e.,  up to a time $t_{max}=\gamma t_r -t_0>0$ when a maximum $\langle y \rangle_{max}=y_0 \exp( t_0/t_r - \gamma) [\gamma t_r/t_0)]^{\gamma}$ is reached. From then on, one has $v_H+v<0$, and the particle is increasingly dragged to the origin as the Hookean force becomes larger.  This behavior is shown in Fig.~\ref{Fig:MeanValueHarmonicPL}, where the time evolution of $\langle y \rangle$ given by Eq.~\eqref{ybarPot} is displayed for $t_r=10^{4}$, $t_0=10^3$ and different values of $\gamma$. For this parameter choice, the relaxation is non-monotonic if $\gamma > 1/10$.

The analytical prediction is in excellent agreement with simulation results which are also shown in Fig.~\ref{Fig:MeanValueHarmonicPL}. To obtain the simulation results in this figure and in all the subsequent figures of the present work, a Gaussian jump length pdf
\begin{equation}
\lambda^{*} (y) = \frac{1}{\sqrt{4 \pi \sigma^2}} \exp \left( \frac{- y^2}{4 \sigma^2} \right)
\label{GaussianPDF}
\end{equation}
with $\sigma^2=1/2$ and the exponential waiting time pdf~\eqref{Exponential PDF} with mean value $\tau=1$ have been used. With this parameter choice, the values of $\varepsilon$ and of the diffusion coefficient [respectively obtained from Eqs.~\eqref{defeps} and~\eqref{DiffusionConstant}] are $\varepsilon=1/\sqrt{\pi}$ and $\mathfrak{D}=1/2$.

From Eq.~\eqref{sigmaOU} and the relation $\sigma_y^2 (t) = a^2(t) \sigma_x^2 (t)$, one can also easily calculate an exact expression for the semi-variance of the physical propagator. One obtains
\begin{equation}
\begin{split}
 \sigma_y^2 (t)=  & \mathfrak{D} t_0 \exp \left[ -\frac{2}{t_r} (t+t_0) \right]  ~ \times \\& \left\{ \left( \frac{t+t_0}{t_0} \right)^{2 \gamma} E_{2 \gamma} \left( -\frac{2 t_0}{t_r} \right) - \left( \frac{t+t_0}{t_0} \right) E_{2 \gamma} \left[ -\frac{2}{t_r} (t+t_0) \right] \right\},
 \end{split}
\label{sy2OEpot}
\end{equation}
where $E_{\nu}(z) = \int_1^{\infty} du \exp(- u z) u^{-\nu} $ denotes the Exponential Integral function.
For large times, the semi-variance approaches the asymptotic value $\sigma_y^2 (\infty)=\mathfrak{D} t_r/2$, as in the static case. This result is valid for any value $\gamma$; it is obtained by inserting the asymptotic approximation (cf. Eq.~(5.1.51) in Ref.~\cite{Abra72})
\begin{equation}
E_{\nu}(z) \sim \exp(-z)/z, \qquad z \to \infty,
\end{equation}
into Eq.~\eqref{sy2OEpot} and by subsequently taking the limit $t \to \infty$. In this limit, one obtains for any value of $\gamma$ a stationary propagator that turns out to be the same as in the static case. In other words, with respect to the static case, a power-law growth or contraction only provokes changes in the transient behavior, but does not significantly affect the confinement induced by the harmonic force in the long time limit. The specific value of $\gamma$ (positive or negative) only has an influence on how fast the propagator converges to the stationary profile. This is also manifested by the fact that the Hubble velocity $v_H=\gamma y/(t+t_0)$ tends to zero as $t$ goes to infinity, and so the result for the static case is recovered in this limit.

\begin{figure}[t]
\includegraphics[width=0.49\textwidth]{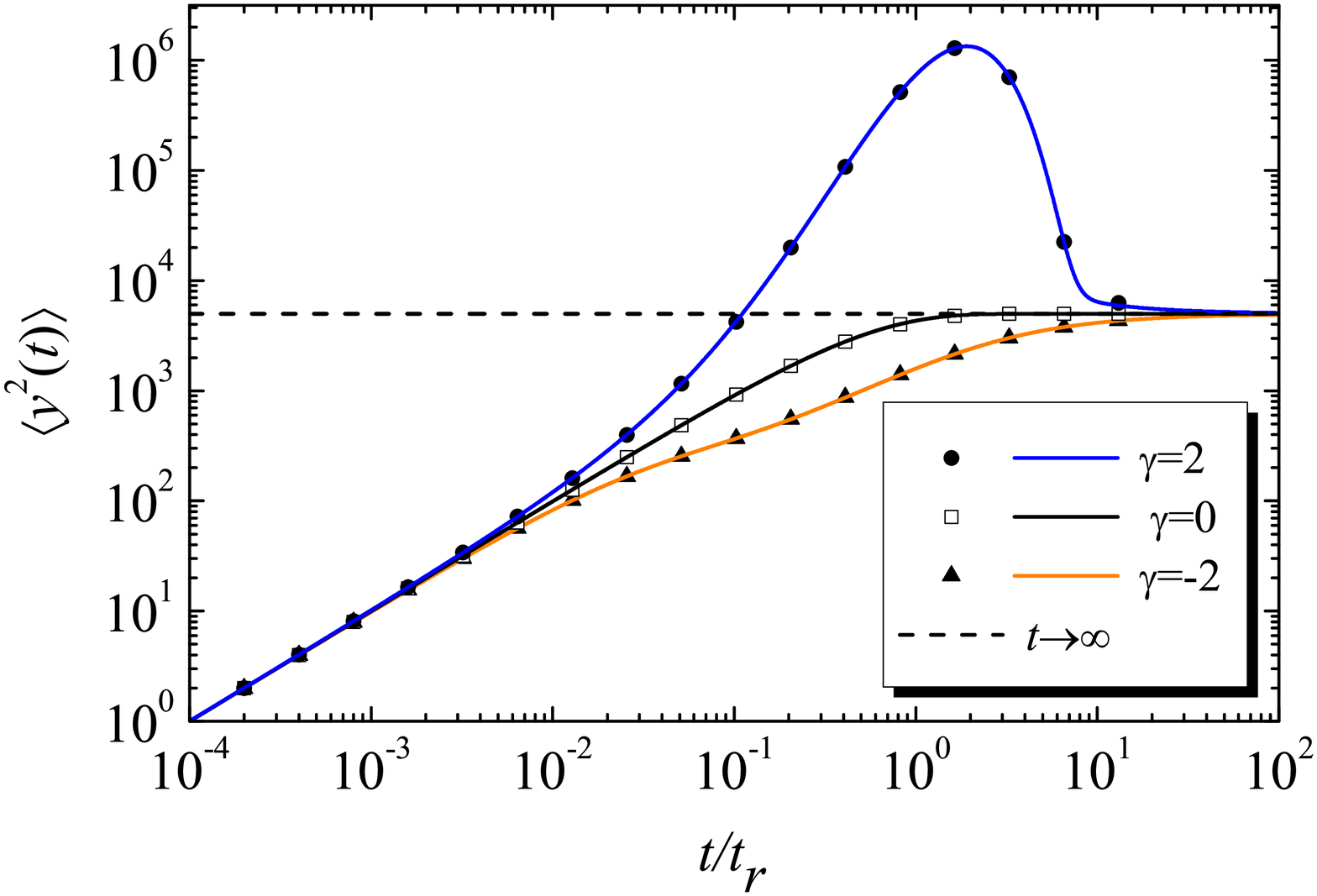}
\caption{Double logarithmic representation of $\langle y^2(t) \rangle = 2 \sigma_y^2$ for $y_0=0$ and a power-law domain growth/contraction [cf. Eq.~\eqref{sy2OEpot}]. The corresponding values of $\gamma$ are, from top to bottom, $\gamma=\{2,0,-2\}$. Further, we have taken $t_0=10^3$, $\mathfrak{D}=1/2$ and $t_r=10^4$. Symbols depict simulation results ($10^6$ runs were performed). The horizontal dashed line represents the asymptotic variance $\mathfrak{D} t_r = 5000.$}
\label{Fig:SecondMomentHarmonicPL}
\end{figure}

The time evolution of the variance $ 2\sigma_y^2 (t)$ is shown in Fig.~\ref{Fig:SecondMomentHarmonicPL} for three different values of $\gamma$ (recall that $\gamma=0$ corresponds to the static case). In the static case, the initial delta-peak first widens, but as soon as the Hookean force starts to become non-negligible, the widening is slowed down; eventually, the variance stabilizes at a fixed value. A power-law contraction of the domain ($ \gamma <0 $) preserves this qualitative behavior, but the transient values of the variance observed before the final value is reached become smaller.

In contrast, for a growing domain ($\gamma>0$), the behavior is non-monotonic; the propagator first widens until a maximum value of $ 2\sigma_y^2 (t)$ is reached, and then it becomes more and more narrow until the stationary profile is reached. Thus, at sufficiently short times, the physical propagator is widened by the combination of diffusion and domain growth; however, beyond a characteristic time, the harmonic force becomes strong enough to limit the dispersion of the particle about the origin induced by the Brownian jumps and by the outward Hubble drift. The characteristic time at which the variance begins to decay can be computed by solving  the equation $ d\sigma_y^2 (t) /dt = 0$ numerically.

\begin{figure}[t]
\includegraphics[width=0.49\textwidth]{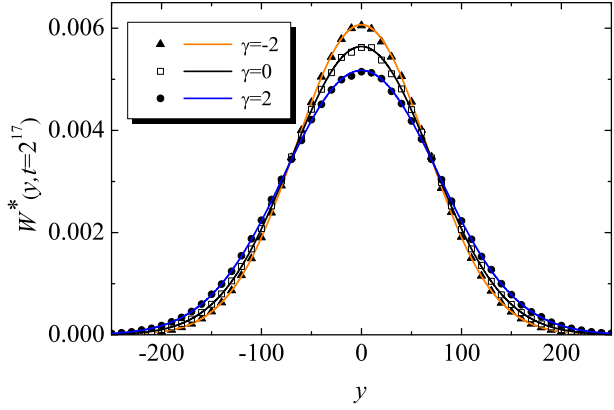}
\caption{Physical propagator $W^{*}(y,t)$ for an OU process on a growing, a static, and a contracting domain at time $t=2^{17}$. In all cases, we have chosen $t_r=10^4$ and $\mathfrak{D}=1/2$. Solid lines depict the analytical propagator, Eq.\eqref{GaussProp} with Eq.\eqref{sy2OEpot}. The growth/contraction rate is dictated by a power law with parameters $t_0=10^3$ and $\gamma=\{-2,0,2\}$ (from top to bottom at $y=0$). The dots, the unfilled triangles and the squares represent the respective simulation results obtained after $10^6$ runs.}
\label{Fig:PHarmonicPL}
\end{figure}

In Fig.~\ref{Fig:PHarmonicPL}, three physical propagators corresponding to three different values of $\gamma=\{-2,0,2\}$  and a common relaxation time $t_r= 10^4$ are depicted for $t=2^{17} \simeq 10^{5.12}$ (the longest simulation time employed in Fig.~\ref{Fig:SecondMomentHarmonicPL}). The value of $t$ has been chosen large enough to ensure that, in the case of a static domain, the obtained propagator is very close to the stationary profile attained for $t\to\infty$ (see
Fig.~\ref{Fig:SecondMomentHarmonicPL}). For $\gamma=2$, one can see that the propagator is slightly more flattened than the stationary one, while for $\gamma=-2$, it is slightly sharper. Note, however, that the typical width of each propagator is practically the same, in agreement with the result of Fig.~\ref{Fig:SecondMomentHarmonicPL} for $t=2^{17}$. As one can see from Fig.~\ref{Fig:PHarmonicPL}, the theoretical curves match very well the random walk simulation results represented by the symbols.

\subsection{Exponential growth}

The case of an exponential scale factor $a(t)=\exp(Ht)$ bears special relevance, since the Langevin equation~\eqref{Leq} tells us that an OU process on a static domain is equivalent to diffusion in an exponentially contracting domain with a suitably chosen value of $H$ (see also Ref.~\cite{Chen2019}). As a result of this equivalence, it is clear that the OU process on an exponentially growing domain may give rise to an interesting competition depending on the values chosen for $\kappa/\xi=t_r$ and for $H$.

More precisely, the semi-variance displays the following $H$-dependence:
\begin{subequations}
\label{Hdepscf}
\begin{itemize}
\item For $H \neq 1/t_r$:
\begin{align}
\sigma_y^2 (t)&= \mathfrak{D} \frac{1 - \exp [2(H-t_r^{-1})t]}{2(t_r^{-1} - H)} .
\label{sy2OUexp1}
\end{align}
\item For $ H = 1/t_r $:
\begin{align}
\sigma_y^2 (t)&= \mathfrak{D} t .
\end{align}
\end{itemize}
\end{subequations}
Note that, for $H\neq 1/t_r$, the semi-variance can be obtained from the result~\eqref{s2yOEestatico} for a static domain by performing the replacement $1/t_r \to 1/t_r-H$. As in the case of the first-order moment, the behavior again depends on the sign of $1/t_r-H$. As we already know, a sufficiently fast growth ($H>1/t_r$) ends up breaking the confinement of the particle. As a result of this, the propagator keeps widening in the limit $t\to\infty$, which is manifested in the divergence of the variance, $\sigma_y^2(t\to\infty)\to \infty$. In contrast, a slow growth ($0<H<1/t_r$) or a contraction ($H<0$) entails convergence of the system to a stationary profile. This behavior is reminiscent of the one observed in the case of a static domain, but in the present case the asymptotic semi-variance is different, $\sigma_y^2 (\infty) = \mathfrak{D}/[ 2(t_r^{-1} - H)]$. Thus, regardless of the smallness of $H$, an exponential growth or an exponential contraction always modifies the width of the stationary propagator obtained in the case of a static domain. Note that this behavior differs from the one observed in the case of a power-law growth. The duration of the transient is also different, since the coefficient $1/t_r$ in the exponential of Eq.~\eqref{s2yOEestatico} is replaced with $1/t_r - H$ in the case of a growing domain [cf. Eq.~\eqref{sy2OUexp1}].

When implemented on a growing domain ($H>0$), the OU process exhibits a dramatic change in behavior. The propagator in physical space is the Gaussian function~\eqref{GaussProp} with mean value $\langle y \rangle=y_0 \exp(t/t_H)/\exp(t/t_r)$ and semivariance given by Eq.~\eqref{s2yOEestatico}, where we have introduced the Hubble time $t_H=1/H$ characterizing the domain growth rate. In contrast with the case of a power-law scale factor, the mean value of the physical coordinate grows exponentially when $t_H<t_r$, or is fixed at the initial position $y_0$ when $t_H=t_r$. Only for $t_r<t_H$ does $\langle y \rangle$ decay to the origin. This interesting behavior reflects a competition between the outward Hubble drift and the Hookean force, whose joint action cancels out for $t_H=t_r$ [in this case, the Hubble velocity $v_H=y/t_H$ and the velocity $v=-y/t_r$ associated with the action of the restoring force compensate each other in Eq.~\eqref{Leq}]. Thus, the particle performs a pure Brownian motion about $y_0$.  One thus concludes that the domain growth is exactly compensated by the harmonic force, implying the equivalence of the latter to the action of a contracting Hubble drift with $|H|=1/t_r$.

\begin{figure}[t]
\includegraphics[width=0.49\textwidth]{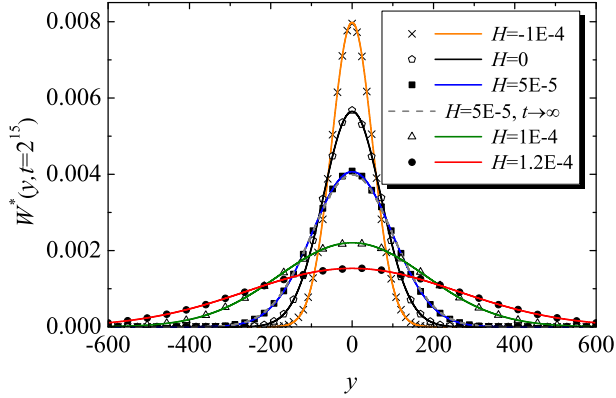}
\caption{Physical propagator $W^{*}(y,t)$ evaluated at time $t=2^{15}$ for a harmonic potential with $t_r=10^4$, a diffusion constant $\mathfrak{D}=1/2$ and an exponential domain growth with different values of $H$. Solid lines correspond to the analytical propagator [Eq.~\eqref{GaussProp} with Eq.~\eqref{Hdepscf}]. We have taken $H=\{-10^{-4}, 0, 5 \times 10^{-5}, 10^{-4}, 1.2 \times 10^{-4} \}$, from top to bottom at $y=0$. Symbols depict simulation results ($10^6$ runs). The dashed line represents the asymptotic profile for $H=5 \times 10^{-5}$.}
\label{Fig:ExponentialHarmonic}
\end{figure}

Displayed in Fig.~\ref{Fig:ExponentialHarmonic} are results for $W^{*}(y,t)$ obtained both from theory and simulations. Profiles corresponding to the initial condition  $W^{*}(y,0)=\delta(y)$ and computed for $t=2^{15} \simeq 10^{4.52}$ are plotted for a fixed value of $t_r=10^4$ and different values of $H$. Depending on the value of $H$, the subsequent evolution of the propagator will be different. In the cases where $H \leq 0$, the depicted propagators are indistinguishable from the stationary one; for the case  $0 < H <1/t_r$, the propagator is close to the stationary one, but can still be distinguished from the latter. This comes as no surprise since, the larger $H$, the slower the decay to the final value [cf. Eq.~\eqref{s2yOEestatico}]. In contrast, when $H \geq 1/t_r$, the propagators widens all the time. Of course, the larger $H$, the wider the corresponding physical propagator.

\section{Subdiffusive OU process on a growing domain}
\label{Sec:OU_SubDiff}

We now turn our attention to the subdiffusive case $\alpha<1$.
In this case, the evolution equation~\eqref{FPE_Harmonic} contains the fractional derivative $~{_0}\mathcal{D}_t^{1-\alpha}$, which arises from the long-time limit of the CTRW process \cite{Metzler00}.  This complicates extraordinarily the task of finding analytical solutions for Eq.~\eqref{FPE_Harmonic}, and one has to resort to numerical approaches. Fortunately, one can obtain explicit expressions for the comoving moments and for the physical moments, whence key insights about the underlying physics can be gained.

As already done for the $\alpha=1$ case, it is convenient to introduce a characteristic relaxation time. To this end, we generalize the definition of $t_r$ as follows:
\begin{equation}
\label{trSub}
t_r=t_r(\alpha)=\left( \frac{\xi_{\alpha}}{\kappa} \right)^{\frac{1}{\alpha}}, \qquad 0<\alpha \le 1.
\end{equation}
The value for the normal diffusive case is recovered in the limit $\alpha\to 1$.

In order to obtain the differential equation governing the evolution of the comoving $m$-th order moment, we follow the standard procedure, i.e., cross-multiplication of Eq.~\eqref{FPE_Harmonic} with $x^m$ and subsequent integration over the domain of the spatial domain. One then obtains a (descending) hierarchy of differential equations:
\begin{equation}
\frac{d \langle x^m \rangle}{dt} = \frac{ m (m-1) \mathfrak{D}_{\alpha} }{a^2} {_0}\mathcal{D}_t^{1-\alpha} \langle x^{m-2} \rangle -\frac{m}{t_r^\alpha} ~ {_0}\mathcal{D}_t^{1-\alpha} \langle x^m \rangle,
\label{avxm_Harm}
\end{equation}
where the definition~\eqref{trSub} has been used. The result~\eqref{avxm_Harm} comes as no surprise, since such hierarchies are typical of diffusion problems. Correspondingly, in physical space one has:
\begin{equation}
\frac{d \langle y^m \rangle}{dt} = m (m-1) \mathfrak{D}_{\alpha} a^{m-2} {_0}\mathcal{D}_t^{1-\alpha} \left[ \frac{\langle y^{m-2} \rangle}{a^{m-2}} \right]  - \frac{m}{t_r^\alpha} a^m {_0}\mathcal{D}_t^{1-\alpha} \left[ \frac{\langle y^m  \rangle }{a^m } \right] + m \frac{\dot{a}}{a} \langle y^m \rangle.
\label{avym_Harm}
\end{equation}
The above equations remain valid for $\alpha=1$, in which case the fractional derivatives are replaced with the identity operator, and a set of ordinary differential equations is obtained.

As in the Brownian case, we are interested in the specific cases of a power-law scale factor and of an exponential scale factor. However, before addressing these two specific cases, we will give some general results for arbitrary $a(t)$.

\subsection{Results for arbitrary scale factor}

Our subsequent discussion will focus on the first- and second-order moments ($m=1,2$). To this end, we shall take
Eqs.~\eqref{avxm_Harm} and~\eqref{avym_Harm} as a starting point.

\subsubsection{First-order moment}
\label{secFirstmoment}

Setting $m=1$ in Eq.~\eqref{avxm_Harm}, one gets
\begin{equation}
\frac{d \langle x \rangle}{dt} =  -\frac{1}{t_r^\alpha} ~ {_0}\mathcal{D}_t^{1-\alpha} \langle x \rangle,
\end{equation}
whose solution is
\begin{equation}
\langle x(t) \rangle = x_0 \,\mathrm{E}_{\alpha,1} \left( - (t/t_r)^{\alpha} \right),
\label{1stcovmom}
\end{equation}
where $\mathrm{E}_{\alpha,1}$ denotes a Mittag-Leffler function \cite{Podlubny1999, Mainardi00}. For $\alpha=1$, this function takes the form of an exponential, and one recovers Eq.~\eqref{averageOU}. Note also that the first-order moment in the comoving coordinate takes the same form as the first-order moment for the case of a static domain \cite{Metzler1999PRE}.

The physical first-order moment $\langle y(t) \rangle = a(t) \langle x(t) \rangle $ will either blow up or decay to the origin depending on whether the scale factor grows faster than the Mittag-Leffler function or not. For negative large arguments, the asymptotic behavior of the Mittag-Leffler function is \cite{Podlubny1999}
\begin{equation}
\mathrm{E}_{\alpha,\beta} \left(- z \right) \sim \frac{1}{z \Gamma(\beta-\alpha)}, \quad z\to \infty,\quad \alpha\neq \beta
\label{asyMLf}
\end{equation}
a result which will be useful to study the long-time behavior of $\langle y(t) \rangle$ for specific forms of the scale factor.

\subsubsection{Second-order moment}

Without loss of generality, we will hereafter assume that $y_0>0$. Although one can consider moments of arbitrary order, the substantial differences between the Brownian case and the subdiffusive case already manifest themselves at the level of the second-order moment. For $m=2$, Eq.~\eqref{avxm_Harm} takes the form
\begin{equation}
\frac{d \langle x^2 \rangle}{dt} = \frac{2\mathfrak{D}_{\alpha}}{a^2} \frac{t^{\alpha-1}}{\Gamma(\alpha)} - \frac{2}{t_r^\alpha} \,{_0}\mathcal{D}_t^{1-\alpha} \langle x^2 \rangle.
\label{avx2_Harm}
\end{equation}
Its counterpart in physical space is
\begin{equation}
\frac{d \langle y^2 \rangle }{dt} = 2\mathfrak{D}_{\alpha} \frac{t^{\alpha-1}}{\Gamma(\alpha)} - \frac{2}{t_r^\alpha} \,{_0}\mathcal{D}_t^{1-\alpha} \left[ \frac{ \langle y^2 \rangle}{a^2} \right] + 2 \frac{\dot{a}}{a} \langle y^2 \rangle .
\label{avy2_Harm}
\end{equation}
In order to analyze the case of a growing domain, let us first recall the main results for the case of a static domain, which is recovered by taking $a(t)=1$ in Eqs.~\eqref{avx2_Harm} and~\eqref{avy2_Harm}. A straightforward way of studying the long-time behavior amounts to first setting $a(t)=1$ in Eq.~\eqref{avy2_Harm} and then taking the Laplace transform of the resulting equation. This yields
\begin{equation}
\langle y^2(s) \rangle = \frac{y_0^2}{s+ (2/t_r^\alpha) s^{1-\alpha}} + \frac{2 \mathfrak{D}_{\alpha}}{s^{1+\alpha}+(2/t_r^\alpha)s}.
\label{x2sHarmonicAlphaH0}
\end{equation}
Taking into account that (Eq. (1.80) of Ref.~\cite{Podlubny1999})
\begin{equation}
\label{LaptbetaE}
\mathcal{L} \left[
t^{\beta-1} \mathrm{E}_{\alpha,\beta}(-a t^\alpha)
\right]
=\frac{1}{s^\beta+as^{\beta-\alpha}},
\end{equation}
one has, from Eq.~\eqref{x2sHarmonicAlphaH0},
\begin{equation}
\label{y2GralEstatico}
\langle y^2(t)\rangle=
y_0^2 \mathrm{E}_{\alpha,1} \left( -2 (t/t_r)^{\alpha}\right)
+
2 \mathfrak{D}_{\alpha} t^{\alpha} \mathrm{E}_{\alpha,1+\alpha} \left( -2 (t/t_r)^{\alpha} \right).
\end{equation}
Then, from Eq.~\eqref{asyMLf}, one finds
that the second-order moment will tend to a constant value in the long-time limit, $\langle y^2(\infty) \rangle = \mathfrak{D}_{\alpha} t_r^\alpha $. As in the case $\alpha=1$, this simply reflects the fact that a balance between diffusive spreading and the confining effect of the restoring force is established (even though the spreading is subdiffusive in the present case).

A procedure similar to the one described above also works for the case of a growing domain, that is, when $a(t)$ is a monotonically growing function in time. From Eq.~\eqref{avx2_Harm}, one finds the Laplace-transformed second-order moment in comoving space:
\begin{equation}
\langle x^2(s) \rangle = \frac{x_0^2}{s +(2/t_r^\alpha) s^{1-\alpha}} + \frac{2 \frac{\mathfrak{D}_{\alpha}}{\Gamma(\alpha)} \mathcal{L} \left[ \frac{t^{\alpha-1}}{a^2(t)} \right]}{s + (2/t_r^\alpha) s^{1-\alpha} } .
\label{x2sHarm_at}
\end{equation}
The first term on the rhs does not depend on the scale factor and yields the same Mittag-Leffler decay as in the static case. Consequently, the contribution of this term to $\langle y^2(t) \rangle$ will be $y_0^2 a^2(t) \mathrm{E}_{\alpha,1} \left( - (t/t_r)^{\alpha} \right)$. As one can see, this contribution can either increase or decrease in time depending on the chosen scale factor. In the next subsections, we will study the joint effect of the first and the second term in Eq.~\eqref{x2sHarm_at} for the special cases of a power-law and of an exponential domain growth.

\subsection{Power-law growth}
\label{Subsec:plg}

\subsubsection{First-order moment}

From Eq.~\eqref{1stcovmom}, one finds the exact expression for the physical first-order moment, i.e.,
\begin{equation}
\langle y(t) \rangle = y_0 \left( \frac{t+t_0}{t_0} \right)^{\gamma} \mathrm{E}_{\alpha,1} \left( -(t/t_r)^{\alpha} \right),
\label{avy_Harm_PL}
\end{equation}
whence the long-time behavior
\begin{equation}
\langle y(t) \rangle \sim \frac{y_0 t_r^\alpha}{\Gamma(1-\alpha) t_0^{\gamma}} t^{\gamma-\alpha}
\label{longtimebeh}
\end{equation}
follows. Thus, for $\alpha<\gamma$, the first-order moment diverges as $t^{\gamma -\alpha}$, whereas for $\alpha>\gamma$, it decays to the origin according to the same power law; finally, when $\alpha=\gamma$, $\langle y(t) \rangle$ displays a plateau in the long-time limit. Beyond this asymptotic behavior, the exact solution~\eqref{avy_Harm_PL} reveals interesting transient effects, reflecting the subtle interplay between the domain growth, the harmonic force, and the diffusive transport.

To start with, note that $\langle y(t) \rangle$ always decays for short times because of the behavior of its time derivative, given by the expression
\begin{equation}
\frac{d \langle y \rangle}{dt} = y_0 \frac{\gamma}{t_0} \left( \frac{t+t_0}{t_0} \right)^{\gamma-1} \mathrm{E}_{\alpha,1} \left( - (t/t_r)^{\alpha} \right) - y_0 t_r^{-\alpha} \left( \frac{t+t_0}{t_0} \right)^{\gamma} t^{\alpha-1} \mathrm{E}_{\alpha,\alpha} \left( -(t/t_r)^{\alpha} \right).
\end{equation}
Indeed, when $\alpha<1$, the second term on the rhs diverges to $-\infty$ as $t \to 0$.

In the case $\gamma>0$, one can distinguish several regimes for the behavior at intermediate times depending on the values of the relaxation time $t_r$ and of a typical time $t_E$ after which the expansion of the domain can be considered to play a relevant role. This typical time can e.g.~be defined as the time at which an arbitrary segment of the domain has doubled its length, $a(t_E)=2$. For the case of power-law growth, this types depends on both characteristic parameters, $t_E=t_E(t_0,\gamma)$. Analogously, in the case of a contracting domain ($\gamma<0$), one can define a typical time $t_C(t_0,\gamma)$ at which the length of a segment is one half of the initial length, $a(t_C)=1/2$.

\begin{figure}
 {\includegraphics[width=0.5\textwidth]{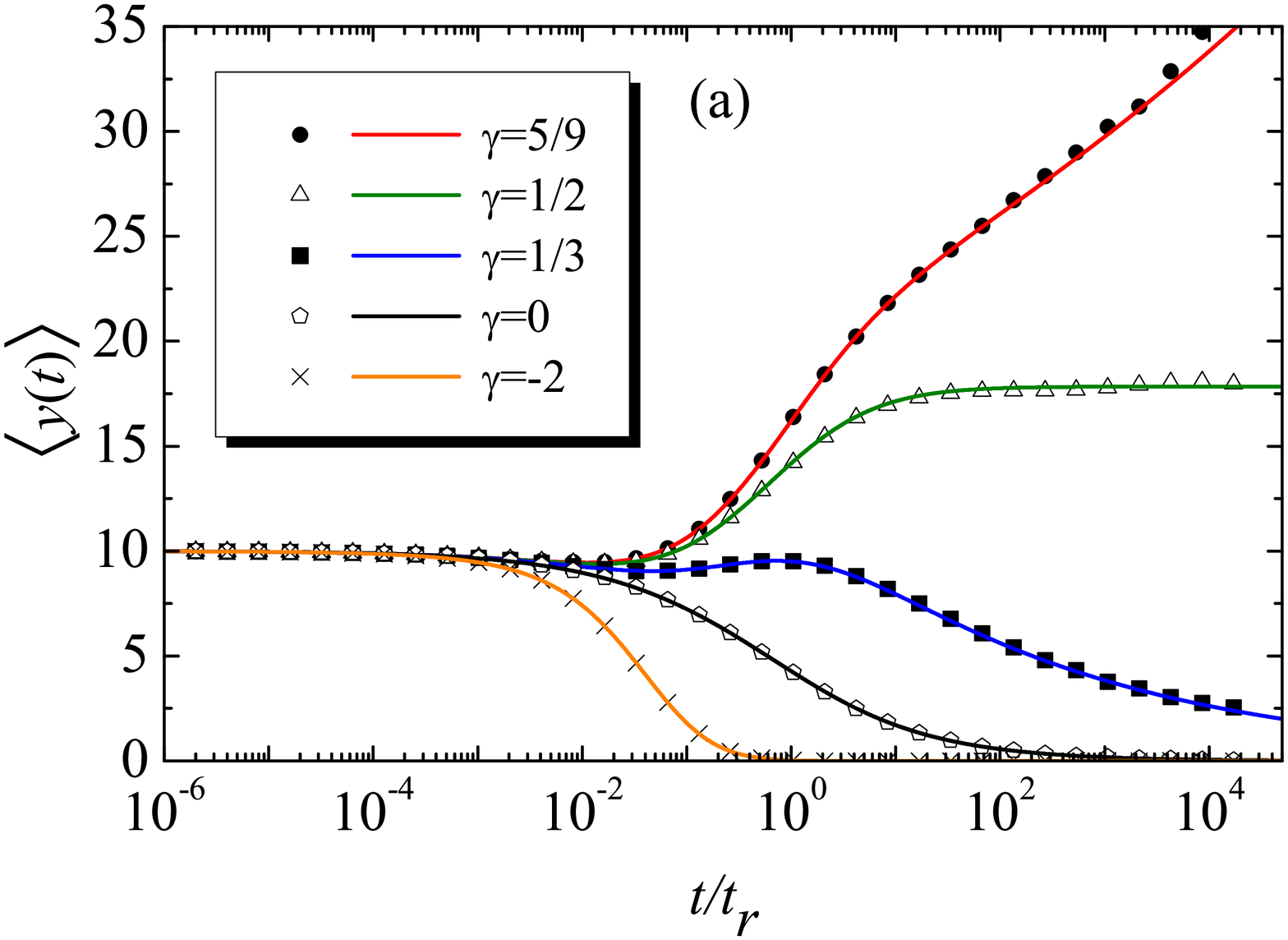}}
 {\includegraphics[width=0.5\textwidth]{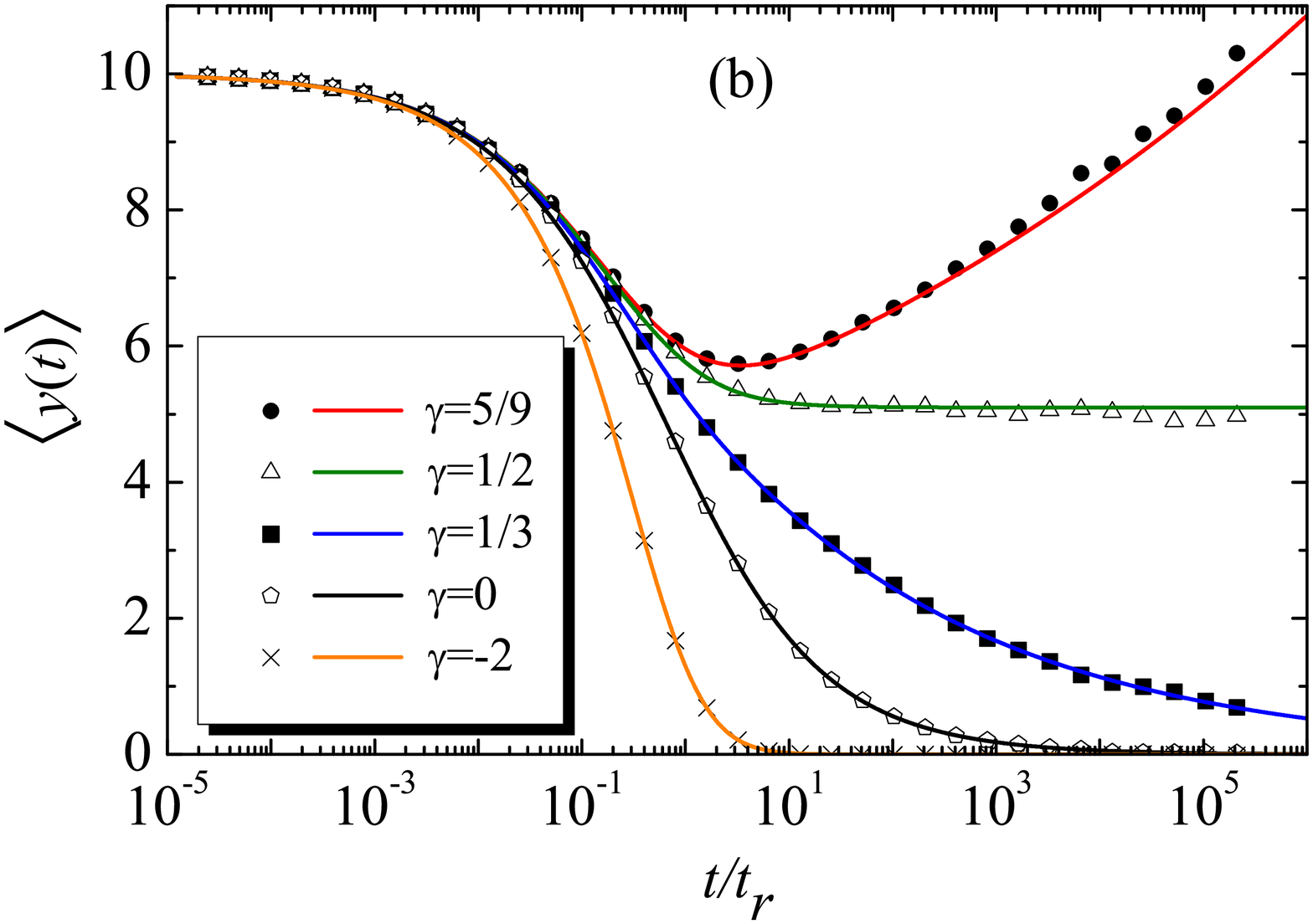}}
\caption{Lin-log plot of the time evolution of $\langle y(t) \rangle$ for a subdiffusive OU process on a domain subject to  a power-law growth/contraction. For the cases of growing domains, one has $t_E<t_r$ in panel (a) and $t_E>t_r$ in panel (b) . The starting point is $y_0=10$. For both panels, we have taken $\alpha=1/2$, $\mathfrak{D}_{\alpha}=1/2$ and $t_0=10^5$. We have chosen $\kappa/\xi_\alpha=10^{-3}$  $(t_r=10^6)$ in panel (a), and $\kappa/\xi_\alpha=3.5\times 10^{-3}$  $(t_r=8.16\times 10^4)$ in panel (b). The solid lines represent the theoretical prediction given by Eq.~\eqref{avy_Harm_PL}, whereas the symbols represent numerical results obtained from $10^6$ simulations of the random walk. The values of $t_E$ for the cases of a growing domain ($\gamma=5/9, 1/2,1/3$) are $2.48\times 10^5, 3\times 10^5, 7\times 10^5$, whereas the value of $t_C$ for the case of a shrinking domain ($\gamma=-2$) is  $4.14\times 10^4$.}
\label{Fig:y1_ParetoHarmonicPL}
\end{figure}

For $t_E \ll t_r$, Fig.~\ref{Fig:y1_ParetoHarmonicPL}(a), the domain expansion begins to play a relevant role comparatively early, implying that $\langle y(t) \rangle$ starts to increase at a time around $t_E$, before finally entering the asymptotic regime described by Eq.~\eqref{longtimebeh}. Note, however, that when $\alpha<\gamma$, $\langle y(t) \rangle$ may still exhibit a local minimum at a time close to $t_E$, i.e., before its pre-asymptotic time growth (see the curve for $\gamma=5/9$ in Fig.~\ref{Fig:y1_ParetoHarmonicPL}). For the specific case $\alpha=\gamma$, note also the onset of the asymptotic plateau predicted by Eq.~\eqref{longtimebeh}.

In the opposite situation where $t_E \gg t_r$, ~\ref{Fig:y1_ParetoHarmonicPL}(b), the intermediate regime vanishes, and so three different behaviors are obtained. For $\alpha > \gamma$, $\langle y(t) \rangle$ becomes a strictly decreasing function; for $\alpha <\gamma$, it exhibits a local minimum; finally, for $\alpha =\gamma$ it tends to a constant value from above.

Turning now to the case of a static domain ($\gamma=0$) and of a shrinking domain ($\gamma<0$), at sufficiently long times the spreading effect of diffusion is overcome by the confinement induced by the restoring force, which is enhanced by the domain contraction in the latter case. As a result of this, a slow decay of $\langle y(t) \rangle$ is observed at sufficiently short times; this slow decay is followed by a faster decay at longer times. This qualitative behavior does not depend on the particular choice of $t_0$ and $t_r$.

\subsubsection{Second-order moment}

\begin{figure}[t]
\includegraphics[width=0.49\textwidth]{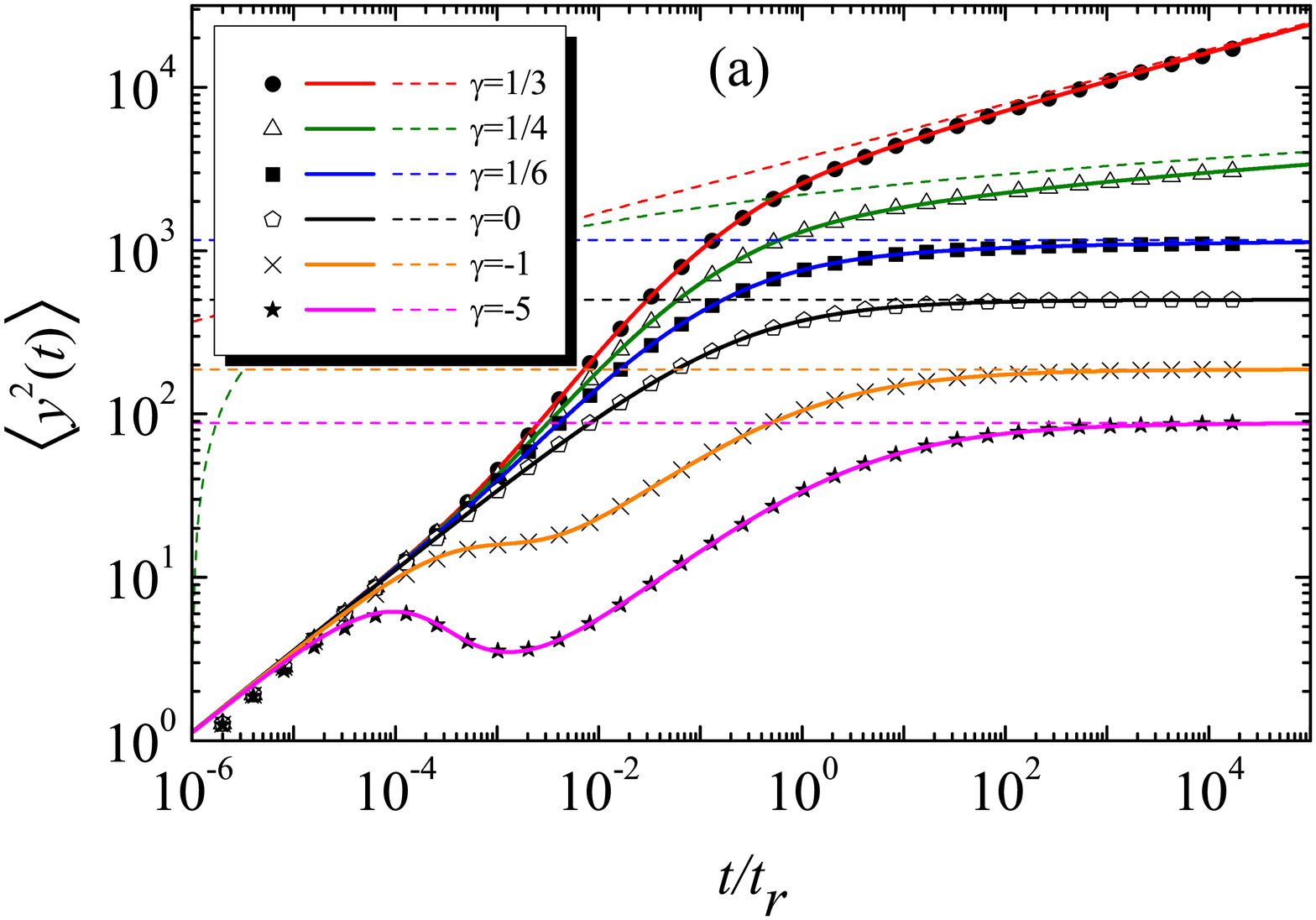}
\includegraphics[width=0.49\textwidth]{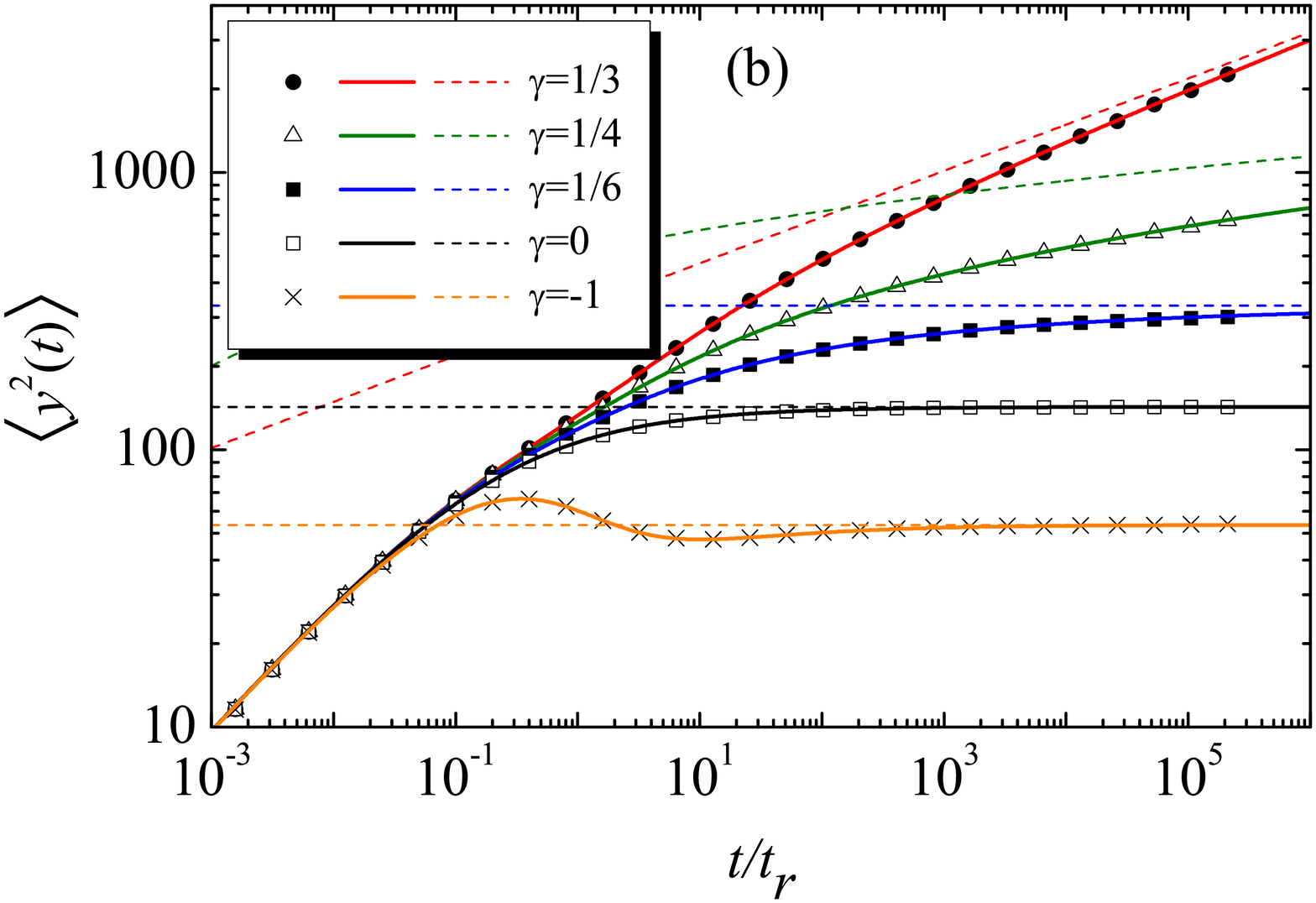}
\caption{Double-logarithmic representation of $\langle y^2(t) \rangle$ for subdiffusive particles on a domain
subject to a power law domain growth/contraction with $t_E, t_C \ll t_r$ [panel (a)] and $t_E, t_C \gg t_r$ [panel (b)]. We have set $\alpha=1/2$, $\mathfrak{D}_{\alpha}=1/2$, $t_r=10^6$ and $y_0=0$. Further, we have used $t_0=10^3$ in panel (a) and  $t_0=10^5$ in panel (b).  The values of $\gamma$ for the cases of domain growth are 1/3,1/4, and 1/6, respectively leading to $t_E=7 \times 10^3, 1.5 \times 10^4$ and $6.3 \times 10^4$ in panel (a) and to $t_E=7 \times 10^5, 1.5 \times 10^6$, and $6.3 \times 10^6$ in panel (b).  For contracting domains, the value $\gamma=-1$ yields $t_C=10^3$ in panel (a) and $t_C=10^5$ in panel (b), whereas for $\gamma=-5$, one has $t_C=148.70$ in panel (a). Solid lines have been computed from a numerical inversion of Eq.~\eqref{x2sHarm_PL}. The different symbols correspond to simulation results. Dashed lines represent the asymptotic long-time behavior given by Eq.~\eqref{y2subcases}. }
\label{Fig:SecondMomentParetoHarmonicPL}
\end{figure}

For a power-law scale factor $a(t)=(1+t/t_0)^\gamma$, Eq.~\eqref{x2sHarm_at} becomes
\begin{equation}
\langle x^2(s) \rangle = \frac{x_0^2 + 2 \frac{\mathfrak{D}_{\alpha}}{\Gamma(\alpha)} t_0^{\alpha} U(\alpha,1+\alpha-2\gamma, s t_0) }{s + (2/t_r^\alpha) s^{1-\alpha} } ,
\label{x2sHarm_PL}
\end{equation}
where $U$ denotes Tricomi's confluent hypergeometric function. The behavior of this function for $s \to 0$ can be found in Ref. \cite{Abra72} (see Eqs. 13.5.6-13.5.12 therein). For $x_0=0$, one has
\begin{equation}
\langle x^2(s) \rangle \sim\mathfrak{D}_\alpha t_r^\alpha \left\{ \begin{array}{lcc}
t_0^{2\gamma} \frac{\Gamma(\alpha-2\gamma)}{\Gamma(\alpha)} s^{2\gamma-1}  &  \text{if} & \gamma<\alpha/2 , \\
\\ -\frac{t_0^{\alpha}}{\Gamma(\alpha)} s^{\alpha-1} \log (s t_0) &  \text{if} & \gamma=\alpha/2 , \\
 \\ t_0^{\alpha} \frac{\Gamma(2\gamma - \alpha)}{\Gamma(2\gamma)} s^{\alpha-1}  &  \text{if}  & \gamma>\alpha/2.
\end{array} \right.
\end{equation}
The long-time behavior of the comoving second-order moment is obtained with the help of a Tauberian theorem \cite{Hughes95}:
\begin{equation}
\langle x^2(t) \rangle \sim  \mathfrak{D}_\alpha t_r^\alpha \left\{ \begin{array}{lcc}
 \frac{\Gamma(\alpha-2\gamma)}{\Gamma(\alpha) \Gamma(1-2\gamma)} \left(\frac{t}{t_0} \right)^{-2\gamma}  &  \text{if} & \gamma<\alpha/2, \\
\\ \frac{\sin(\pi \alpha)}{\pi} \left(\frac{t}{t_0} \right)^{-\alpha} \log(t) &  \text{if} &  \gamma=\alpha/2, \\
 \\  \frac{\Gamma(2\gamma - \alpha)}{\Gamma(2\gamma) \Gamma(1-\alpha)} \left(\frac{t}{t_0} \right)^{-\alpha}  &  \text{if}  &  \gamma>\alpha/2.
\end{array} \right.
\end{equation}
Consequently, in physical space one has
\begin{equation}
\langle y^2(t) \rangle \sim \mathfrak{D}_\alpha t_r^\alpha \left\{ \begin{array}{lcc}
 \frac{\Gamma(\alpha-2\gamma)}{\Gamma(\alpha) \Gamma(1-2\gamma)} &  \text{if} & \gamma<\alpha/2, \\
\\ \frac{\sin(\pi \alpha)}{\pi} \log(t) &  \text{if} &  \gamma=\alpha/2, \\
 \\  \frac{\Gamma(2\gamma - \alpha)}{\Gamma(2\gamma) \Gamma(1-\alpha)} \left(\frac{t}{t_0} \right)^{2 \gamma - \alpha}  &  \text{if}  & \gamma>\alpha/2.
\end{array} \right.
\label{y2subcases}
\end{equation}
Thus, three different asymptotic regimes can be distinguished depending on the values of $\alpha$ and $\gamma$. These theoretical results are confirmed by numerical simulations, see Fig.~\ref{Fig:SecondMomentParetoHarmonicPL}. This behavior, which is very different from the one observed in the Brownian case, will be discussed in detail in what follows.

To start with, we note that only a sufficiently slow domain growth ($\gamma<\alpha/2$) can give rise to the onset of a stationary state. Such a steady state results from the trade-off between diffusive spreading (enhanced by the Hubble drift) and the restoring force directed towards the origin (cf. Fig.~\ref{Fig:SecondMomentHarmonicPL} with Fig.~\ref{Fig:SecondMomentParetoHarmonicPL}). In addition, the asymptotic value $\langle y^2(\infty) \rangle$ displays a dependence on $\gamma$ that was absent in the $\alpha=1$ case. In other words, when $\alpha<1$, the signature of the domain growth persists for arbitrarily long times.

Let us now consider the opposite case of a fast domain growth ($\gamma>\alpha/2$). When $\alpha<1$, the distribution of subdiffusive particles becomes increasingly sparse as a result of an increasing interparticle distance. The variance displays an unbounded growth, as opposed to what happens when the particles are Brownian ($\alpha=1$).  Since the change in behavior in the vicinity of $\alpha=1$ is somewhat counterintuitive, the underlying physics must be carefully explained.

For Brownian walkers as well as for subdiffusive walkers, the typical distance traveled by diffusion up to a given time is proportional to the mean number of steps taken.  However, Brownian walkers typically cover a much longer distance than subdiffusive ones; the reason is that the jump rate is constant for a Brownian walker, while for a subdiffusive one, it decreases in time due to the ageing of the CTRW process. Such ageing effects lead to anomalously long waiting times during which the walker is under the sole influence of the Hubble drift, which tends to induce a strong particle separation. The harmonic potential cannot totally counteract this effect, because CTRW particles only \textit{feel} this potential when they take instantaneous jumps. In spite of this, the effect of the restoring force is manifested by the existence of different asymptotic regimes. In the forceless case, one has $\langle y^2(t) \rangle \propto t^{2\gamma}$ at long times, since diffusive transport plays a minor role in comparison with the Hubble drift \cite{LeVotAbadYuste17}. In contrast, when the restoring force is at play, one has a slower spreading $\langle y^2(t) \rangle \propto t^{2 \gamma-\alpha}$. Thus, a power-law domain growth, no matter how fast, is unable to completely suppress the signature of the harmonic potential in the long-time regime.

In the marginal case $\gamma=\alpha/2$, the behavior characteristic of the driftless case is also modified by a prefactor $t^{-\alpha}$. Indeed, as shown in Ref. \cite{LeVotAbadYuste17},  one has $\langle y^2(t) \rangle \propto t^{\alpha} \log(t)$ in the absence of the harmonic force, whereas the growth dictated by Eq.~\eqref{y2subcases} is purely logarithmic.

Finally, note that for $\gamma>\alpha/2$,  the long-time behavior of $\langle y^2(t)\rangle$ depends not only on the time exponent $\gamma$ of the scale factor, but also on $t_0$.

The approach of $\langle y^2(t) \rangle$ to the asymptotic regime is another important point (see Fig.~\ref{Fig:SecondMomentParetoHarmonicPL}). Let us first consider the case $\gamma>0$. Once again, a discussion in terms of $t_E$ and $t_r$ is pertinent. When $t_E \ll t_r$, Fig.~\ref{Fig:SecondMomentParetoHarmonicPL}(a), the time evolution of the physical variance displays three different regimes, namely, a diffusion-controlled early-time regime, an intermediate regime where both diffusion and domain growth contribute significantly to the walker's motion, and a final asymptotic long-time regime.  In a double logarithmic representation, the slope of  $\langle y^2(t) \rangle$ when the crossover from the early-time regime to the intermediate regime takes place can only grow
[see Fig.~\ref{Fig:SecondMomentParetoHarmonicPL}(a)]. In the opposite case $t_E \gg t_r$, one also has an intermediate regime, in this case due to the coexistence of diffusion and a non-negligible harmonic force. Here, the slope of  $\langle y^2(t) \rangle$ decreases when the crossover from the early-time regime to the intermediate regime takes place [see Fig.~\ref{Fig:SecondMomentParetoHarmonicPL}(b)].

However, the most interesting scenario appears for power-law contractions ($\gamma<0$), see the corresponding curves in Figs.~\ref{Fig:SecondMomentParetoHarmonicPL}(a) and ~\ref{Fig:SecondMomentParetoHarmonicPL}(b). For Brownian particles, $\langle y^2(t) \rangle$ is a non-decreasing function of time, but in the present case $0<\alpha<1$ it may exhibit an interesting transient oscillation.  The $\gamma<0$ cases depicted in Fig.~\ref{Fig:SecondMomentParetoHarmonicPL} (a) correspond to the situation $t_C\ll t_r$. After an initial growth of $\langle y^2 (t) \rangle$ due to subdiffusive spreading, the joint contribution of the force field and of the inward Hubble drift tends to drive the particle towards the origin, and eventually a stationary physical propagator with finite width settles.

\begin{figure}[t]
\includegraphics[width=0.49\textwidth]{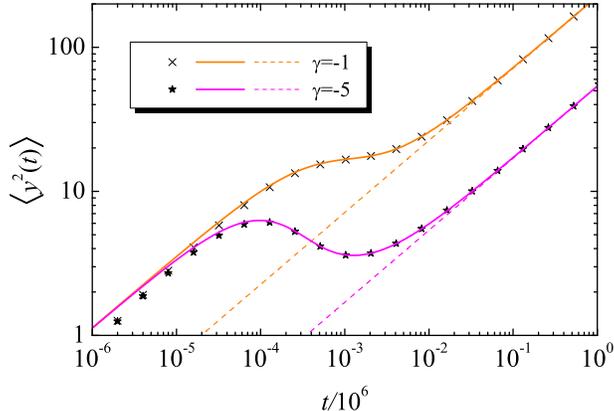}
\caption{Double-logarithmic representation of $\langle y^2(t) \rangle$ for subdiffusive particles on a power-law growing domain in the absence of the restoring force ($\kappa=0)$. Represented are the curves for $\gamma=-1,-5$. The remaining parameters have been chosen as in Fig.~\ref{Fig:SecondMomentParetoHarmonicPL}. Dashed lines represent the asymptotic behavior for large times $\langle y^2 (t) \rangle \sim 2 \mathfrak{D}_{\alpha} t^{\alpha} / [\Gamma(\alpha) (\alpha-2\gamma)]$ (see Ref.~\cite{LeVotAbadYuste17}). In the horizontal axis, the time variable $t$ has been divided by the value of $t_r$ chosen in Fig.~\ref{Fig:SecondMomentParetoHarmonicPL}(a) in order to facilitate the comparison between both figures.}
\label{Fig:y2_k0_PLContraction.eps}
\end{figure}

In order to explain the above phenomenon, it is convenient to first examine the force-free case, since a transient oscillation may already arise in this case for a sufficiently large value of $|\gamma|$. From Ref. \cite{LeVotAbadYuste17}, it is known that the spreading effect of a subdiffusive CTRW prevails over the confining effect of a Hubble drift arising from a power-law contraction. Thus, the variance diverges in the long-time limit. However, for large enough $|\gamma|$, the power-law contraction may result in a transient decrease of the variance before the final time growth characteristic of the long-time asymptotics is eventually attained (see Fig.~\ref{Fig:y2_k0_PLContraction.eps}).

When a harmonic potential is incorporated into the above picture, the confining effect of the Hubble drift is enhanced. If the elastic constant $\kappa$ is sufficiently small, one has $t_r \gg t_C$. In this case, if an oscillation is observed well before $t_r$, it will be preserved, and only the long-time asymptotics will be affected by the restoring force, implying that the variance will no longer diverge as in the force-free case, but rather tend to a constant value [see curves for $\gamma=-1,-5$ in Fig.~\ref{Fig:SecondMomentParetoHarmonicPL}(a)]. This reflects the fact that, at times $t \sim t_r$,  the coordinate $y$ takes values which are sufficiently large to ensure that the increased harmonic force effectively counterbalances the effect of subdiffusive spreading.

In the case $t_C \gg t_r$,  a transient oscillation can still be seen [cf. the $\gamma=-1$ curve
in Fig.~\ref{Fig:SecondMomentParetoHarmonicPL}(b)]. At times that are roughly between $t_r$ and $t_C$, the dynamics essentially corresponds to the case of a static domain, i.e., the growth of the variance is counterbalanced by the harmonic force, and the variance tends to a finite value. At longer times, the contraction tends to transiently drive the particles towards the origin, but this is eventually unsufficient to counterbalance the effect of diffusive spreading, and the variance goes to a constant value at times
$t\gg t_C$. This actually happens for arbitrarily small values of $|\gamma|$ (provided that the condition
$t_C \gg t_r$ holds).  When the Hubble drift becomes significant, the particles have a weaker tendency to spread than in the case $t_C \ll t_r$; therefore, the recovery of the amplitude is not so strong in this case, and the oscillation is somewhat ``dampened''.

\subsection{Exponential growth}
\subsubsection{First-order moment}

From Eq.~\eqref{1stcovmom}, in the case of an exponentially growing (or decreasing) scale factor the physical first-order moment is
\begin{equation}
\langle y (t) \rangle = y_0 \exp(H t) \mathrm{E}_{\alpha,1} \left( - (t/t_r)^{\alpha} \right) .
\end{equation}
As in the power-law case, $\langle y (t) \rangle$ is a decreasing function at short times because the derivative of this function,
\begin{equation}
\frac{d \langle y \rangle}{dt} = y_0 \exp(H t) \left[ H \mathrm{E}_{\alpha,1} \left( - (t/t_r)^{\alpha} \right) - (t/t_r)^{\alpha-1} \mathrm{E}_{\alpha,\alpha} \left( - (t/t_r)^{\alpha} \right) \right],
\end{equation} is negative and divergent as $t\to 0$.

\begin{figure}[t]
\includegraphics[width=0.49\textwidth]{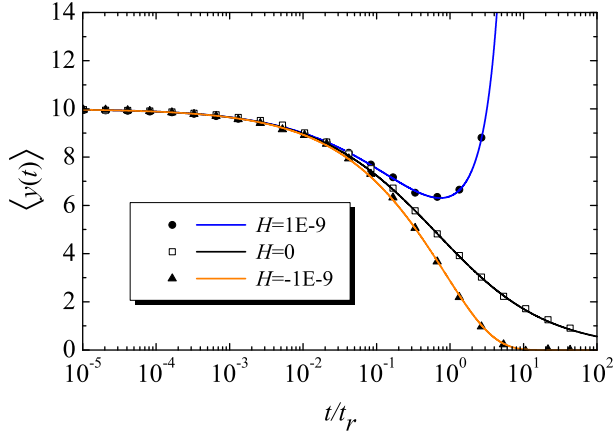}
\caption{Physical first-order moment of a subdiffusive particle initially located at $y_0=10$ on an exponential expanding/shrinking domain. The other parameter values are $\alpha=1/2$, $\mathfrak{D}_{\alpha}=1/2$, $t_r^{-1}=2.5 \times 10^{-9}$, and $H=\{10^{-9}, 0, -10^{-9} \}$, from top to bottom. Symbols represent simulation results obtained from $10^6$ runs.}
\label{Fig:y1_ParetoHarmonicExp}
\end{figure}

In contrast, the long-time behavior depends once again on the sign of $H$. For $H>0$, $\langle y(t) \rangle$ blows up, since the slow asymptotic decay of the Mittag-Leffler function is not able to compensate the fast growth of the exponential. Because of the change in sign of the time derivative, one must have a minimum given by the condition $\left. d\langle y (t) \rangle / dt = 0 \right|_{t=t_{min}}$ (see Fig.~\ref{Fig:y1_ParetoHarmonicExp}).  The corresponding time $t_{min}$ can be computed numerically. It is expected to be of the order of $t_H$, which is roughly the time scale for which the Hubble drift starts to play an important role.

For the case $H<0$ of a contracting domain, the first-order moment decays to zero faster than in the case of a static domain. One actually expects that for times $t \gtrsim 1/|H|$, $\langle y(t) \rangle$ will already be very close to zero.

\subsubsection{Second-order moment}

An exponential scale factor $a(t)=\exp(Ht)$ destroys the stationary state resulting from the tradeoff between subdiffusive spreading and the restoring force. For $H>0$, the variance grows without bound, whereas an exponential contraction ($H<0$) induces a very strong particle localization about the origin. We will first address the case of exponential growth ($H>0$), since the procedure employed to obtain the long-time behavior of the physical second-order moment is the same as in the case of a power-law domain growth (cf. subsection~\ref{Subsec:plg}).

According to Eq.~\eqref{x2sHarm_at}, the Laplace-transformed comoving second-order moment reads as follows:
\begin{equation}
\langle x^2 (s) \rangle = \frac{x_0^2 + 2 \mathfrak{D}_{\alpha} (s+2H)^{-\alpha}}{s + (2/t_r^\alpha) s^{1-\alpha}}.
\end{equation}
For $s\to 0$, one has
\begin{equation}
\langle x^2 (s) \rangle \sim \frac{x_0^2 + 2 \mathfrak{D}_{\alpha} (2H)^{-\alpha}}{s +(2/t_r^\alpha) s^{1-\alpha}},
\label{x2s_Harm_ExpH>0}
\end{equation}
whence the long-time behavior
\begin{equation}
\langle x^2 (t) \rangle \sim \left( x_0^2 + 2^{1-\alpha} \mathfrak{D}_{\alpha} H^{-\alpha}\right) \text{E}_{\alpha, 1} \left(-2 (t/t_r)^{\alpha} \right), \quad H>0,
\end{equation}
follows by virtue of a Tauberian theorem. Correspondingly, the asymptotic growth of the physical second-order moment
is given by the expression
\begin{equation}
\langle y^2 (t) \rangle = a^2(t) \langle x^2 (t) \rangle \sim \left( x_0^2 + 2^{1-\alpha} \mathfrak{D}_{\alpha} H^{-\alpha}\right) \text{E}_{\alpha, 1} \left( (t/t_r)^{\alpha} \right) \exp(2Ht), \quad H>0.
\label{varexpgrowth}
\end{equation}
As we can see from the above equation and from Eq.~\eqref{asyMLf}, regardless of the values of $t_H$ and $t_r$, an exponential growth always overcomes the confining effect of the harmonic potential; subdiffusive particles are thus able to spread further and further. The behavior is therefore different from the one observed in the Brownian case $\alpha=1$, characterized by either confinement (for $t_H=1/H> t_r$) or unlimited dispersal (for $t_H=1/H<t_r$). The behavior also differs from the case of a power-law domain growth, where (for subdiffusive particles) the influence of the harmonic potential is present for arbitrarily long times. Here, the signature of the force appears in the argument of the Mittag-Leffler function via $t_r$, but the long-time behavior is clearly dominated by the exponential $\exp(2Ht)$.  In contrast, recall that, for a power-law growth, Eq.~\eqref{y2subcases} states that the variance goes to a constant value for $\gamma<\alpha/2$, increases logarithmically for $\gamma=\alpha/2$, or increases as a power-law for $\gamma>\alpha/2$ (as already mentioned, in the latter case the asymptotic time growth $\propto t^{2\gamma}$ observed on a static domain is dampened with a prefactor $t^{-\alpha}$ induced by the restoring force).

\begin{figure}[t]
\includegraphics[width=0.49\textwidth]{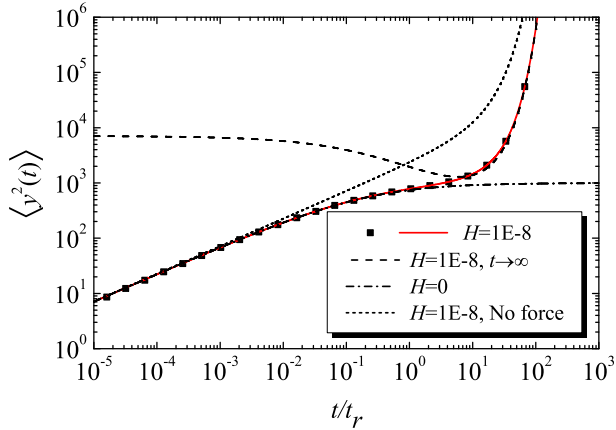}
\caption{Double logarithmic plot of the time evolution of the physical second-order moment for a subdiffusive particle initially located at $y_0=0$ on an exponentially growing domain. We have set $\alpha=1/2$, $\mathfrak{D}_{\alpha}=1/2$, $t_r =4\times 10^6$, and $H=10^{-8}$. The squares represent the numerical results obtained after $10^6$ simulation runs. The solid line is the theoretical curve. The dashed line corresponds to the long-time behavior as given by Eq.~\eqref{varexpgrowth}. The two remaining curves correspond to the cases with either no force (short-dashed line) or no domain growth (dash-dotted line).}
\label{Fig:ExponentialParetoHarmonic}
\end{figure}

Displayed in Fig.~\ref{Fig:ExponentialParetoHarmonic} is the physical second-order moment for $H=10^{-8}$ and $t_r =4\times 10^6$. An excellent agreement between the theoretical curve and the simulation result is found. The theoretical curve stems from the numerical computation of the inverse Laplace transform of Eq.~\eqref{x2s_Harm_ExpH>0}. As one can see, this curve overlaps with the one for the static case until a time of the order of $1/H$, and blows up at longer times. At very long times, the theoretical curve tends to the curve for the force-free case.

\subsubsection{Second-order moment for an exponential contraction}

Let us now focus on the case of exponential contraction, corresponding to a scale factor $a(t)=\exp(Ht)$ with $H<0$. In the absence of forces, random walkers obeying the heavy-tailed waiting time distribution~\eqref{heavytailwtd} will be strongly localized at the origin at sufficiently long times (in fact, the physical propagator tends quickly to a delta function $\delta(y)$, a behavior that was termed ``Big Crunch'' in Refs.~\cite{LeVotAbadYuste17, LeVotYuste2018}). A harmonic potential with a minimum at the origin will enhance the Big Crunch effect, in the sense that the strong narrowing of the pdf will take place at earlier times than in the force-free case. As a result of this, all the moments (and in particular the second-order one) will quickly go to zero.

\begin{figure}[t]
\centering
\includegraphics[width=0.49\textwidth]{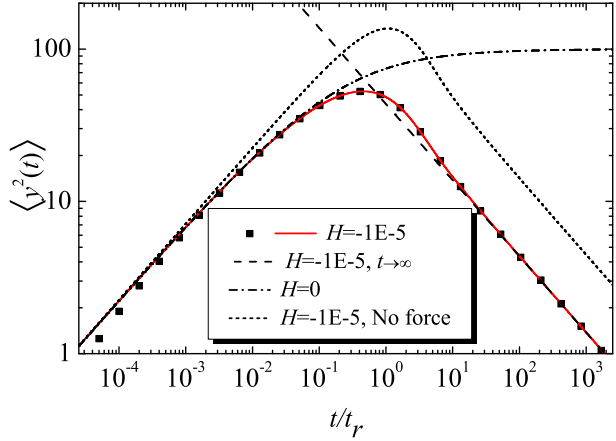}
\caption{Double logarithmic representation of $\langle y^2(t) \rangle$ for $y_0=0$, $\alpha=1/2$, $t_r=4\times 10^4$, $\mathfrak{D}_{\alpha}=1/2$, and $H=-10^{-5}$. The squares represent simulation results obtained after $10^6$ runs. The solid curve corresponds to the numerical inversion of~\eqref{Phys2M_Harmonic_Subdiffusive_Exponential_Laplace}. The dashed line corresponds to the asymptotic long-time behavior, Eq.~\eqref{Phys2M_Harmonic_Subdiffusive_Exponential_As1}. The other lines correspond to the static case $H=0$, Eq.~\eqref{y2GralEstatico} (dash-dotted line) and to the force-free case $\kappa=0$ (short-dashed line).}
\label{Fig:BigCrunchHarmonic}
\end{figure}

For our subsequent analysis, it is convenient to directly work in physical coordinates. For exponential expansions or contractions, it is very easy to find the Laplace-transformed second-order moment from Eq.~\eqref{x2s_Harm_ExpH>0} by means of the shift theorem. One gets
\begin{equation}
\langle y^2 (s) \rangle = \langle x^2 (s-2H) \rangle =\frac{y_0^2+2\mathfrak{D}_{\alpha} s^{-\alpha}}{s - 2H + (2/t_r^\alpha) (s - 2H)^{1-\alpha}} .
\label{Phys2M_Harmonic_Subdiffusive_Exponential_Laplace}
\end{equation}
For small $s$, one has
\begin{equation}
\langle y^2 (s \to 0) \rangle \sim \frac{\mathfrak{D}_{\alpha}}{t_r^{-\alpha} (-2H)^{1-\alpha} -H} s^{-\alpha}, \quad H<0.
\end{equation}
The corresponding Tauberian theorem then provides the long-time behavior:
\begin{equation}
\langle y^2 (t \to \infty) \rangle \sim \frac{\mathfrak{D}_{\alpha}}{t_r^{-\alpha} (-2H)^{1-\alpha} -H} \frac{t^{\alpha-1}}{\Gamma(\alpha)} \quad H<0.
\label{Phys2M_Harmonic_Subdiffusive_Exponential_As1}
\end{equation}
As expected, one obtains an asymptotic inverse-power decay to the origin with exponent $\alpha-1$. In Fig.~\ref{Fig:BigCrunchHarmonic}, the time evolution of the second-order moment in the presence of a Hookean force associated with a relaxation time $t_r=4\times 10^4$ (solid curve) is compared with the force-free case (short-dashed curve) for a Hubble parameter $H=-10^{-5}$. As one can see, the turning point (maximum) is slightly shifted to an earlier time by the harmonic force. This enhanced Big Crunch effect highlights the strength of an exponential contraction, as opposed to a combination of the harmonic potential with a weaker, power-law contraction. Recall that, in the latter case, the effect of subdiffusive spreading is not totally overcome by the domain contraction, and this results in a non-vanishing value of the asymptotic variance $\langle y^2 (\infty) \rangle$ (cf. Fig.~\ref{Fig:SecondMomentParetoHarmonicPL}).

\subsection{Numerical solution}

In contrast with the Brownian case, finding exact expressions for the propagator of subdiffusive CTRWs on a growing domain turns out to be a very difficult task. In the case of a static domain, one can obtain the propagator of a subdiffusive particle in the presence of an external force field from the corresponding propagator for a Brownian particle  \cite{Metzler00}, but this method does not work for a growing domain. The source of the difficulties stems from the fact that, in the present case, knowing the probability of having taken a certain number of steps $n$ up to a given time $t$ is not enough; one must know the probabilities that the $n$ steps of the random walk were taken at specific times $<t$ \cite{LeVotYuste2018}. In spite of this drawback, Eq.~\eqref{FPE_Harmonic} can be integrated numerically by means of the fractional Crank-Nicolson method of Ref.~\cite{Yuste2006}. This method is an unconditionally stable finite-difference method where the space-time region of integration  is discretized by a mesh spacing of size $\Delta x$ and time steps of size $\Delta t$. Shortly we will see that the agreement of the numerical integration procedure with numerical simulations is excellent, except in the immediate vicinity of the points where the propagator is non-differentiable.

\begin{figure}[t]
\centering
\includegraphics[width=0.49\textwidth]{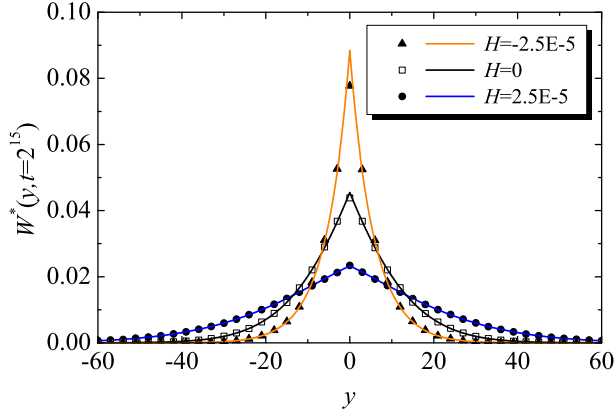}
\caption{Physical propagators evaluated at time $t=2^{15}$ for exponential scale factors with $H= \left\{ -2.5\times 10^{-5}, ~0, ~2.5\times 10^{-5} \right\}$, from top to bottom at $y=0$. The solid lines corresponding to the numerical solutions of Eq.~\eqref{FPE_Harmonic}; symbols depict results from random walk simulations ($10^6$ runs). In all cases, the subdiffusive particles are characterized by the parameters $\alpha=1/2$, $\mathfrak{D}_\alpha = 1/2$, and $t_r= 10^6$. }
\label{Fig:PHarmonic_Subdiffusive_Exp}
\end{figure}

Fig.~\ref{Fig:PHarmonic_Subdiffusive_Exp} shows the physical propagators corresponding to different values of $H$ for a subdiffusive particle initially located at $y_0=0$ and subsequently subjected to the action of the harmonic potential. Obviously, in this case all the propagators are symmetric with zero-mean. As already seen from the results obtained for the moments, the domain growth/contraction only modifies the specific value of the typical width, but it does not affect the qualitative long-time behavior (recall that for $H<0$ the propagator tends to $\delta(y)$, whereas for $H>0$ it never stops getting flat).

\begin{figure}[t]
\centering
\includegraphics[width=0.49\textwidth]{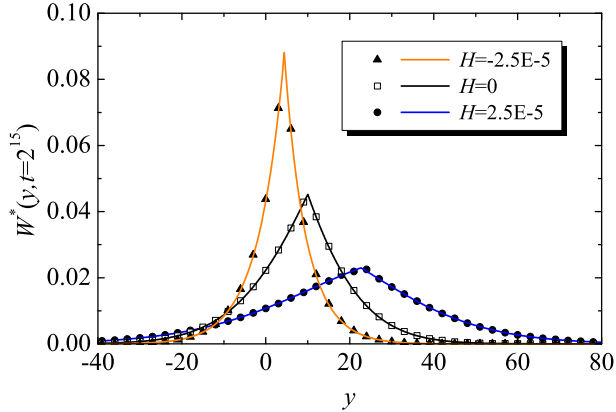}
\caption{Physical propagators evaluated at time $t=2^{15}$ for the initial condition $W^*(y,0)=\delta(y-10)$. Three different exponential expansions are considered: $H= \left\{ -2.5\times 10^{-5}, ~0, ~2.5\times 10^{-5} \right\}$. The solid lines depict the numerical solution of Eq.~\eqref{FPE_Harmonic}, whereas symbols represent results from random walk simulations ($10^6$ runs). In all cases, $\alpha=1/2$ and $\mathfrak{D}_\alpha = 1/2$, and $t_r=10^6$.}
\label{Fig:PHarmonic_Subdiffusive_y10}
\end{figure}

In contrast, when $y_0 \neq 0$, the propagators are non-symmetric with respect to $y=0$, as shown in Fig.~\ref{Fig:PHarmonic_Subdiffusive_y10} for $y_0=10$. In this figure, one can see that the Hubble drift and the restoring force shift the cusp of the propagator from $y_0$ to a different position in the course of time. Note, however, that the propagator remains at all times non-differentiable at $y=y_0 a(t)$. It should also be noted that the numerical integration procedure yields spurious values in the vicinity of $y_0 a(t)$ because of the sharp, delta-peaked initial condition (which is approximated by the function $1/\Delta x$ for $x_0-\Delta x/2<x<x_0+\Delta x/2$ and 0 otherwise), and because of the finiteness of the integration step. While this unwanted effect can be minimized by reducing the size of the time step, this procedure quickly becomes very costly in terms of CPU time. Therefore, we have replaced the spurious numerical results at $a(t) y_0$ with the linear extrapolation of the numerical values of $W(x,t)$  [or, equivalently, $W^*(y,t)$] calculated from the two nearby mesh points situated to the left, say, of $a(t) y_0$. In Figs.~\ref{Fig:PHarmonic_Subdiffusive_Exp} and~\ref{Fig:PHarmonic_Subdiffusive_y10}, we have used $\Delta x = 0.1$ and $\Delta t =0.1$. The size of the integration domain was always chosen large enough to ensure that the finite boundary effects were negligible.

\section{Conclusions}
\label{Sec_Conclusions}

In this paper we have studied diffusion processes in a harmonic potential (i.e., Ornstein-
Uhlenbeck processes) on uniformly expanding/contracting domains. We have found that the impact of a domain growth process on the properties of diffusing particles may be very different depending on whether these are Brownian or subdiffusive.
More precisely, the asymptotic equilibrium between the random force responsible for diffusive spreading and the restoring force may or may not break down depending on the type of diffusion process and on how fast the domain expands. For instance, in the case of power-law domain growth, the propagator eventually reaches the same equilibrium state as on a static domain. When the particle's motion is governed by a subdiffusive CTRW, the system may still tend to an equilibrium state (albeit one that is different from the stationary state attained on a static domain); however, it is also possible that the long-time behavior is essentially controlled by the domain growth when the latter is sufficiently strong. In the case of an exponential scale factor, the long-time behavior of a subdiffusive CTRW  is always determined by the domain growth/contraction. However, for Brownian particles, this only happens if the Hubble parameter $H$ exceeds a certain positive threshold value; otherwise, the system tends to a stationary propagator, albeit different from the one attained on a static domain.

The extreme sensitivity of CTRW particles to the domain growth can be ascribed to ageing; the diffusive spreading of such particles proceeds more and more slowly in the course of time due to a decrease in the jump rate. As a result of this, the Hubble drift, which acts even when the particles do not jump, becomes increasingly relevant as $t$ grows; eventually, it ends up playing a key role in the dynamics.

Another important difference between Brownian and subdiffusive CTRW particles discussed here concerns the functional form of the propagator. In the case of a growing domain Brownian particles on a growing/shrinking domain still follow a Gaussian distribution, but for CTRW particles the functional form of the propagator changes with respect to the one corresponding to the static case. This holds true even in the absence of a force. This feature is a signature of the memory effects induced by the subdiffusive CTRW model.

We close by noting that the present work can be generalized in many ways, e.g. by considering other types of subdiffusive processes, such as fractional Brownian motion with Hurst exponent $<1/2$. Because of the lack of ageing in this case, the phenomenology observed in the CTRW case is expected to change drastically. Finally, the study of superdiffusive processes such as L\'evy flights in the above context is also of interest.

\section{Acknowledgments}

This work was partially funded by the Spanish Agencia Estatal de Investigaci\'on through Grants No. FIS2016-76359-P (partially financed by FEDER funds) (S.~B.~Y. and E.~A.) and by the Junta de Extremadura through Grant No. GR18079 (S.~B.~Y. and E.~A.). F.~L.~V. acknowledges financial support from the Junta de Extremadura through Grant No. PD16010 (partially financed by FSE funds).

\section*{Appendix: Some remarks on the simulation algorithm}

In the simulations, the relation between the force and the bias given by Eq.~\eqref{FrictionConstant} will be taken into account. To simulate the CTRW process underlying Eq.~\eqref{EDADME_Force}, we have used the following effective force:
\begin{equation}
F^*(y) =  \left\{ \begin{array}{lcc}
            -\kappa y &  &  \text{for} \quad |y| \leq d_c, \\
              -\kappa d_c   & & \text{for} \quad  |y|>d_c,
             \end{array}
   \right.
\end{equation}
with $d_c \equiv 2\xi_{\alpha} \varepsilon \sigma / (\tau^{\alpha} \kappa )$. The simulation results obtained with this force at a given time $t$ should practically be the same as the results obtained from Eq.~\eqref{FPE_Harmonic}, provided that $t$ is sufficiently large for the walker to enter the diffusive regime, while the walker remains within a region of radius $d_c$ with a probability close to one. If the latter condition is not fulfilled, the diffusive description is not appropriate; the reason is that the walker's motion becomes deterministic, since the probability of the walker to jump towards the origin is one as soon as $|y|\ge d_c$ [c.f. Eq.~\eqref{FrictionConstant}]. In the case of a sufficiently fast domain growth, the drift induced by the latter pulls the walker away from the origin, implying that the typical distance traveled by the walker will exceed $d_c$; the walker is then practically forced to take the next jump in the opposite direction. The parameters in the simulations have been chosen in such a way that the distance traveled by the walkers is almost always smaller than $d_c$, and so the diffusive description holds.

\end{document}